\documentclass[12pt,a4paper]{article}
\usepackage[active]{srcltx}
\usepackage{graphicx}
\usepackage{bbm}
\usepackage{amsmath}
\usepackage{amssymb}

\renewcommand{\includegraphics}[1]{}

\def \ha {{1 \over 2}}
\def \td {\tilde}
\def \ci{\cite}

\def\foot{\footnote}

\def \ci {\cite}
\def \ov {\over}
\def \bp {\begin{pmatrix}}  \def \epm {\end{pmatrix}}
\def \ha {{\textstyle{1 \ov 2}}}
\def \bi{\bibitem}
\def \la {\label}
\def \l {\lambda}
\def\foot{\footnote}

\def \sql {{\sqrt \l}}
\def \adss {$AdS_5 \times S^5$}
\newcommand{\rf}[1]{(\ref{#1})}
\def \ci {\cite}
\def   \adsc  {AdS$_4\times \IP^3$}
\def   \adss  {AdS$_5\times $S$^5$}
\newcommand{\order}[1]{ {\cal O}\left(\frac{1}{#1}\right)}

\def \J {{\cal J}} \def \S {{\cal S}} \def \E {{\cal E}}

\def \no {\nonumber}

\newcommand{\ellJ}{\mathbb{J}}

\def\IZ{\mathbb{Z}}
\def\IP{\mathbb{CP}}

\def\id{\protect{{1 \kern-.28em {\rm l}}}}

\def\be{\begin{eqnarray}}
\def\ee{\end{eqnarray}}
\def \s {\sigma}

\def\nn{\nonumber}
\def \vp {\varphi}

\makeatletter
\renewcommand\section{\@startsection {section}{1}{\z@}%
                                   {-3.5ex \@plus -1ex \@minus -.2ex}%
                                   {2.3ex \@plus.2ex}%
                                   {\normalfont\large\bfseries}}
\renewcommand\subsection{\@startsection{subsection}{2}{\z@}%
                                   {-3.25ex\@plus -1ex \@minus -.2ex}%
                                   {1.5ex \@plus .2ex}%
                                   {\normalfont\normalsize\bfseries}}
\makeatother

\def\Tr{{\rm Tr}}

\def\AdSS{AdS$_5\times$S$^5$}
\def\AdSP{AdS$_4\times\IP^3$}

\def \rmJ {{J}}

\begin{document}

\thispagestyle{empty}
\begin{flushright}\footnotesize
\texttt{AEI-2008-077}
\texttt{ Imperial-TP-AT-2008-5 }
\vspace{0.8cm}
\end{flushright}

\renewcommand{\thefootnote}{\fnsymbol{footnote}}
\setcounter{footnote}{0}

\begin{center}
{\Large\textbf{\mathversion{bold}
Quantum spinning strings in   AdS$_4\times \IP^3$:\\
 testing the Bethe Ansatz proposal 
}\par}

\vspace{1.5cm}

\textrm{Tristan McLoughlin$^1$, Radu Roiban$^2$ and Arkady A. Tseytlin$^{3,}$\footnote{Also 
at Lebedev Institute Moscow} } \vspace{8mm}

\textit{$^{1}$
Max-Planck-Institut f\"ur Gravitationsphysik\\
Albert-Einstein-Institut\\
Am M\"uhlenberg 1, 14476 Potsdam, Germany}\\
\vspace{3mm}

\textit{$^{2}$
Department of Physics, Pennsylvania State University\\
University Park, PA 16802, USA}\\
 \vspace{3mm}

 \textit{$^{3}$
   The Blackett Laboratory, Imperial College,
London SW7 2AZ, U.K.       }\\
 \vspace{3mm}

\textit{$^{1}$}\texttt{tristan.mcloughlin@aei.mpg.de},\ 
\textit{$^{2}$}\texttt{radu@phys.psu.edu},\ 
\textit{$^{3}$}\texttt{tseytlin@ic.ac.uk} 

\par\vspace{1cm}

\textbf{Abstract} \vspace{5mm}

\begin{minipage}{14cm}

Recently, an asymptotic Bethe Ansatz that is claimed to  describe anomalous
dimensions of ``long'' operators in the planar ${\cal N}=6$
supersymmetric three-dimensional Chern-Simons-matter theory dual to
quantum superstrings in AdS$_4\times\IP^3$ was proposed. It initially
passed a few consistency checks but subsequent direct comparison to
one-loop string-theory computations created some controversy. Here we
suggest a
resolution by pointing out that, contrary to the initial assumption
based on the algebraic curve considerations, the central interpolating
function $h(\lambda)$ entering the BMN or magnon dispersion relation
receives a non-zero one-loop correction in the natural string-theory
computational scheme. We consider a basic example which has already
played a key role in the $AdS_5 \times S^5$ case: a rigid circular
string stretched in both AdS$_4$ and along an $S^1$ of $\IP^3$ and carrying   two
spins. Computing the leading one-loop quantum correction to its energy
allows us to fix the constant one-loop term in $h(\lambda)$ and also
to suggest how one may establish a correspondence with the Bethe
Ansatz proposal, including the non-trivial one-loop phase factor.
We discuss some problems which remain 
in trying to match a
part of world-sheet contributions 
(sensitive to compactness of the worldsheet space-like direction) and
 their Bethe Ansatz  counterparts.

\end{minipage}

\end{center}

\def \ha {{\textstyle{1\ov 2}}}

\vspace{0.5cm}

\newpage
\setcounter{page}{1}
\renewcommand{\thefootnote}{\arabic{footnote}}
\setcounter{footnote}{0}

\tableofcontents


\newpage
\section{Introduction}

The duality \ci{ABJM} between planar ${\cal N}=6$ supersymmetric
three-dimensional Chern-Simons-matter theory and free type IIA
superstring theory in \adsc\ (AdS/CFT$_3$ for short) has attracted
much attention recently.  This is for a good reason, as both the
perturbative gauge theory and the dual free string theory appear to be
integrable (as was partially verified at two-loop level in gauge
theory -- namely in the scalar sector \ci{mz} -- and at the classical
level in string theory \ci{AF,Stef}).  If so, this correspondence may
be providing us with a second example of integrable gauge-string
duality, in addition to the by now well understood canonical one
relating the ${\cal N}=4$ super-Yang-Mills theory (SYM) and the \adss\
superstring (or AdS/CFT$_4$).
   
   Being less than maximally supersymmetric, this new duality is
   useful as it reveals various seemingly obvious assumptions that
   were made (and eventually shown to be correct in the maximally
   supersymmetric context) in the construction of the solution for the
   spectrum of AdS/CFT$_4$ based on the Bethe Ansatz (see \ci{bes} and
   references therein).  Bearing in mind possible future studies of
   less supersymmetric dualities in both three and four dimensions
   this is an important step forward.
   
One crucial change compared to the AdS/CFT$_4$ case is that now the
BMN or magnon dispersion relation is no longer protected and receives
nontrivial corrections both in the weak and strong coupling expansions
\ci{gaiotto,ter} (see also \ci{gv,AhN,GHO,Berenstein:2008dc}). 
For example,  the dispersion relation for the 
``lighter''
magnon and its  BMN limit are given by
\be
\epsilon(p)=\frac{1}{2}\sqrt{1+
16 h^2(\lambda)\sin^2\frac{p}{2}}
~~~~~~ \to ~~~~
\frac{1}{2}\sqrt{1+ 16 \pi^2 h^2 (\lambda) \frac{k^2}{{\rmJ}^2} } \ ,  
\label{dis}
\ee
with
$p=\tfrac{2 \pi k}{\rmJ}$ in the BMN limit and where
\be \la{hah}
    h(\l \ll 1 ) = {\l} [1+  c_1 \l^2 + c_2 \l^4 + ...] \ , \ \ \ \ \ \ \ \ 
    h(\l \gg 1 ) = \sqrt{ \l \ov 2 } + a_1   +{ a_2 \ov {\sqrt{\lambda}}} +...  \ . 
\ee
Incorporating this new interpolating function, the authors of \ci{gv}
made a remarkable proposal for the corresponding Bethe Ansatz which
has (somewhat surprisingly at first sight) essentially the same
structure as in the AdS/CFT$_4$ case.  It was suggested in
\ci{gv} that the leading  strong coupling (one-loop in the world sheet theory) 
 correction to $h(\lambda)$ should vanish, i.e. $a_1=0$; this was
 apparently confirmed in \ci{shend} where the fluctuation spectrum
 near the giant magnon solution was computed using the algebraic curve
 technique \ci{dvi,gv1,gv2} 
 (the conjecture also passed
a few other consistency checks see \cite{AhN,Astolfi:2008ji}).

However, the subsequent direct string theory computations 
\ci{mlr, AAB, CK} 
of the one-loop correction to the universal scaling function, i.e. the
coefficient of the $\ln S$ term in the folded spinning string energy
\ci{GKP,FT1}, led to the result that was different from the Bethe
Ansatz prediction of \ci{gv} based on the assumption that $a_1=0$.
 
 It was suggested in \ci{grmik} that this disagreement was due to
 different regularizations used, or rather to different ways of
 combining fluctuation frequencies in the calculation of the one-loop
 correction to the string energy.  The proposed prescription, argued
 to be intrinsic to the algebraic curve description of the classical
 string solutions in the Bethe Ansatz context, favored the $a_1=0$
 choice.

 While the string theory sigma model is manifestly one-loop finite in
 the ultraviolet, separate terms in one-loop corrections contain
 logarithmic divergences. Hence results obtained by regularizing
 separate terms in different ways, e.g. using different cutoffs, 
  may
 differ by finite terms (for an example, see \ci{tirz-tse-recent}). On
 general grounds, however, in the string theory calculation one should
 regularize the world-sheet action or the path integral; any
 acceptable regularization should be independent of the fine structure
 of the spectrum of fluctuations around a specific solution\footnote{
    Ideally, the classical solution should be constructed
    in the presence of the regulator.
 }
 and should
 preserve the basic (global and local) symmetries of the theory.
 Within the class of {\it acceptable} world-sheet regularizations all
 choices should be equivalent.

Our aim here will be to provide a resolution to the apparent
contradiction between the world-sheet \ci{mlr, AAB, CK} and the Bethe
Ansatz \ci{gv,grmik} calculations while staying within a natural and
consistent world-sheet regularization scheme. We will be led to the
conclusion that, in this context, the coefficient $a_1$ in equation
\rf{hah} has a non-zero value
\be 
a_1 = -  { \ln 2  \ov 2 \pi}  ~~.
\la{hh} 
\ee 
Using this value in the Bethe Ansatz prescription of \ci{gv} restores
the agreement between the string theory result and the Bethe Ansatz
result for the one-loop term in the universal scaling function. 
%
%
%

A non-zero value for the constant term $a_1$ may be accounted for by a
redefinition of the 't~Hooft 
coupling\foot{
Here we consider the
coupling as it appears on the string worldsheet and thus have in mind
redefinition of the form $\tfrac{1}{\sqrt{\lambda'}}=\tfrac{1}{\sqrt{\lambda}}
-a_1\tfrac{1}{\lambda}+\dots$. 
It is not {\it a priori} clear that such a redefinition will be
consistent with a similar weak coupling redefinition of the form
$\lambda'= \lambda + c_1 \lambda^2 + \dots $.},
%
%
suggesting that the world-sheet and
the Bethe Ansatz calculations effectively employ different
regularization schemes.  While anomalous dimensions at renormalization
group fixed points are scheme-independent, for conformal field
theories parameterized by free parameters the scheme dependence may,
in fact, arise as the freedom of redefining these parameters.  Such
may be the case here, in contrast with the world-sheet theory in
\AdSS\ where no such redefinitions appear to be necessary.

A possible way of avoiding such an
ambiguity is to define the coupling constant of the theory in terms of
an observable, e.g.  a particular anomalous dimension. Perhaps a natural
choice for such an observable is the universal scaling function 
$f(\l)$. Eliminating the 't~Hooft coupling in favor of $f$ effectively removes
all scheme ambiguities 
related to  coupling constant redefinitions.
Such a proposal was put forward in
QCD  \cite{physical_coupling} to systematically account for the
scheme dependence in the running of the coupling constant. Since at weak coupling 
$f(g(\l))\sim \l$, the resulting expressions are
necessarily analytic in $f$. This analyticity property holds also 
(despite a different dependence on the 't~Hooft coupling) for the
gauge theory dual of the  world-sheet theory in \AdSP. 

It is worth noting that in the all-loop Bethe Ansatz  proposal
  of \ci{gv} the 't Hooft
coupling appears only through the function $h(\lambda)$ and
consequently if we express all other anomalous dimensions in terms of
the scaling function $f(h(\lambda))$ any  trace of the function $h$ will
be removed,  demonstrating that it is unphysical. However,  this is only
true for the Bethe Ansatz of \ci{gv}; for the perturbative calculation in the
gauge or string theory we must work with $\lambda$ and thus need the
explicit weak or strong coupling expansion of $h(\lambda)$ in whatever
regularization scheme we choose to work in.

\

In addition, below  we will be able to provide a non-trivial test of the
proposal of \ci{gv} by directly computing the one-loop correction to
the energy of the circular $(S,J)$ string \ci{ft2,art,PTT} from the
AdS$_4\times \IP^3$ string theory action and then trying to match the
result with the prediction of the Bethe Ansatz of \ci{gv}.
We will find that the two answers are remarkably similar,  indicating that
the Bethe Ansatz proposal of \ci{gv} may 
indeed
be correct at
strong coupling.
However, few issues remain, warranting a further more systematic study
on the Bethe Ansatz side.

\

The computation of the one-loop correction \ci{PTT,bt,sz} to the
energy of the simplest rigid circular $(S,J)$ string in \adss\ played
a key role in discovering the presence of the one-loop term \ci{bt,hl}
in the phase in the strong-coupling (or ``string'') form of the Bethe
Ansatz \ci{afs}. Our plan here will be  to follow the same logic as in
\ci{PTT,bt}, i.e. carry out the analogous computation in the \adsc\
case and then compare to the Bethe Ansatz prediction.

For later use  let us
define the rescaled coupling constant, $\bar \l$, in terms
 of the 't Hooft coupling, 
$\l$ (equal to ${ N\ov k_{\rm cs}}$, where 
$k_{\rm cs}$ is the level of the Chern-Simons action),
\foot{Since we are interested in the strict 
 `t Hooft limit when $N\to \infty, \   k_{\rm cs}
\to \infty$ with  $\l$ being fixed we can treat $\l$ as a continuous parameter.}
as well as the function ${\bar h}$ as
\be\la{hed}
{\bar\lambda}=2\pi^2 \lambda \ , 
~~~~~~~~~~~~~~~~
{\bar h}({\bar\lambda})= 2 \pi h(\l) ~~.
\ee
The role of ${\bar\lambda}$ is to emphasize the close analogy between
the AdS$_5$ and the AdS$_4$ string-theory expressions.

\bigskip

Let us first recall the story in the \adss\ case. The $(S,J)$ string
solution of \ci{art} has a spiral-like shape, with projection to
$AdS_3$ being a constant radius circle (with winding number $k$), and
projection to $S^5$ -- a big circle (with winding number $m$). The
corresponding spins are, respectively, $S$ and $J$ with the Virasoro
condition implying that $u\equiv {S \ov J} = - {m \ov k}$.  The
classical string energy has the following expansion in large
semiclassical parameters $\S$ and $\J$ with fixed $k$ and fixed $u={\S
\ov \J}$ \ci{art,PTT} ($E_0= \sql
\E(\S,\J,k), \ \S= { S \ov \sql}, \ \J= { J \ov \sql}$, \ $ {\sql \ov
2 \pi} $ is the string tension)
\be 
&& E_0 = S + J   + { \l \ov J} e_1(u,k)  + 
 { \l^2 \ov J^3} e_3(u,k) +  { \l^3 \ov J^5} e_5(u,k) + ...  \ ,
\la{ops} 
\ee
where $e_1= {k^2 \ov 2} u(1+u), \ \ e_3= - {k^4 \ov 8} u(1+u)(1+ 3u
+u^2),$ \ $e_5= { k^6 \ov 16} u(1+u)(1+ 7u +13u^2+ 7u^3 +u^4)$, etc.
In the limit when $u\to 0$ or $S \ll J$ this takes the familiar BMN
form
\be 
\la{bmn}
 E_0 = J + \sqrt{ 1 + {\l k^2\ov J^2}} \ S\ + O(S^2) \ .
\ee 
Computing the one-loop correction $E_1= \E_1(\S,\J,k)$ to the energy gives
\ci{PTT,bt} 
\be 
E_1 = E_1^{\rm even} + E_1^{\rm odd}, \ \ \ \ \ \ E_1^{\rm even} = {
\l \ov J^2} g_2(u,k) + { \l^2 \ov J^4} g_4(u,k) +...\ , \ \ E_1^{\rm
odd}= { \l^{5/2} \ov J^5} g_5(u,k) + ...
\la{lps} 
\ee 
The absence of the $1\ov J$ and $ 1\ov J^3$ terms here implies the
non-renormalization of the BMN-type part of the classical energy
\rf{bmn} which is consistent with the non-renormalization of the BMN
dispersion relation in the \adss\ case. This also suggests that the two
leading ${ \l\ov J}$ and ${ \l^2\ov J^3}$ terms are protected and
their coefficients should directly match the corresponding one-loop
and two-loop perturbative gauge theory results.

 Indeed, the coefficient 
$g_2$ of the  ``even''  $1\ov J^2$ term\foot{Its value is 
$g_2=- \ha M^2 + \sum_{n=1}^\infty [ n \sqrt{ n^2 + 4 M^2 }   - n^2 - 2 M^2],$\ \  
\ $M^2\equiv k^2 u(1+u)$.}
 in \rf{lps} can be reproduced as a leading $1\ov J$ (finite spin
 chain length) correction from the one-loop Bethe Ansatz in the
 $sl(2)$ sector of the ${\cal N}=4$ SYM theory \ci{btz}. An extension
 to higher orders was discussed in \cite{szz}.
 The same should apply to the coefficient of the other analytic even
 ${ \l^2 \ov J^4}$ term -- i.e. it should match the two-loop gauge
 theory result.

At the same time, the presence of the non-analytic in $\l$ and ``odd''
in $1\ov J$ term ${ \l^{5/2} \ov J^5}$ in \rf{lps} (with $g_5= {k^6
\ov 3} u^3(1+u)^3 $) implies that a similar $1\ov J^5$ term in the
classical energy \rf{ops} is not protected so that its coefficient
cannot be directly compared to three-loop result on the gauge theory
side. This resolves the infamous ``three-loop disagreement'' \ci{ss}
and implies \ci{bt} that the corresponding ``string'' Bethe Ansatz
\ci{afs} should be modified to contain a non-trivial one-loop
correction to the phase.\foot{The one-loop term in the S-matrix
dressing phase can be completely determined by including higher order
terms in the expansion of $E_1$ \ci{hl}.}
 
 \bigskip
 
The circular $(S,J)$ string solution in \adsc\ is essentially
the same as that in
\adss, with  the classical energy  having again the form \rf{ops} 
(modulo some numerical factors due 
to the different definition of string tension).  
However,
as we shall find below, the expression for the one-loop correction is
drastically changed: the expansion of $ E_1^{\rm odd}$ in \rf{lps}
starts already with $1\ov J$ and $1\ov J^3$ terms. This implies that
the corresponding leading terms in the classical energy \rf{ops} are
no longer protected.\foot{This is of course not surprising given that
the leading gauge-theory correction here is the two-loop one \ci{mz},
i.e. proportional to $\l^2$, while the leading term in the classical string energy
still scales as $\l$.} Indeed, considering the $S\ll J $ limit,
i.e. comparing to equation \rf{bmn}, these odd one-loop corrections
can be unambiguously interpreted as a one-loop renormalization of the
coefficient of the ${ k^2\ov J^2}$ term under the square root in the
BMN dispersion relation \rf{bmn}, leading to the value of the one-loop
shift in $h(\l)$ given in \rf{hh} (cf. equations \rf{dis}, \rf{bmn}).

Several of the $1\ov J^5$ terms can similarly be interpreted as
arising from the one-loop shift in $h(\l)$; the remaining term happens
to be essentially the same as in \adss\ case, which is in perfect
agreement with the Bethe Ansatz proposal of
\ci{gv} where the S-matrix dressing phase has the same form as in the
AdS/CFT$_4$ case (up to the replacement of $\sqrt{\l}$ by $2 \bar h
(\bar \lambda)=4\pi h(\l)$).

 The even $1\ov J^2$ and $1\ov J^4$ terms do not appear to be  the same as in
 the \adss\ case, but can be formally related to their \adss\ counterparts 
 by restricting the sum over mode
 numbers to odd integers 
%
and making some re-identification of parameters.  While the results of
our computation appear to be in agreement with the general structure
of the Bethe Ansatz of \ci{gv} with $h(\l)$ given by
\rf{hah},\rf{hh} 
there are still  remaining subtle issues related to $1\ov J^{2n}$ terms 
which require further clarification.

 \bigskip

The rest of this paper is organized as follows. In section 2 we 
discuss, following closely the model of \AdSS~\cite{art, PTT},
the structure of the classical circular string solution in \adsc.  In
section 3 we present the spectrum of quadratic fluctuations near this
solution derived directly from the Green-Schwarz superstring action. 
In section 4 we
sum up these frequencies to derive the one-loop correction to the string
energy. We then compare it to the similar expression in the \adss\
case, determining in the process the one-loop term in the $h(\l)$
function and discussing  correspondence with the Bethe Ansatz result
implied by the proposal of \ci{gv}. 
Some computational details and special cases 
are
collected in five appendices.


\section{The circular rotating string in \AdSP}

As was recently pointed out \ci{ABJM}, the closed superstring (type
IIA) background which describes holographically the $U(N)\times U(N)$
${\cal N}=6$ Chern-Simons theory at levels $(k_{\rm cs}, -k_{\rm cs})$
is (we follow the notation of \ci{mlr})
\be
ds^2&=&\frac{R^3}{4k_{\rm cs}}\left(ds_{\rm AdS_4}^2+4ds^2_{\IP^{3}}\right)\ , 
~~~~~~~~~~
e^{2\phi}=\frac{R^3}{k_{\rm cs}^{3}}
\cr
F_{2}&=&k_{\rm cs}\ \ellJ_{\IP^{3}}\ , 
~~~~~~~~~~~~~~
F_{4}=\frac{3}{8}R^3{\rm Vol}_{\rm AdS_4} 
\label{bkgrnd}
\ee
Here 
\be
\label{ads_metric}
ds^2_{{\rm AdS}_4}&=&
-\cosh^2\rho\ dt^2+d\rho^2+\sinh^2\rho 
\left(d\theta^2+\sin^2\theta d\phi^2\right)\ , 
\\
\label{p3_metric}
ds^2_{\IP^{3}}&=& d\zeta_1^2
+\sin^2\zeta_1\left[ d\zeta_2^2+\cos^2\zeta_1
\left(d\tau_1+\sin^2\zeta_2\left(d\tau_2+\sin^2\zeta_3d\tau_3\right)\right)^2
\right. \nn\\
& &\left. 
+ \sin^2\zeta_2\left( d\zeta_3^2+\cos^2\zeta_2
\left(d\tau_2+\sin^2\zeta_3d\tau_3\right)^2
+\sin^2\zeta_3\cos^2 \zeta_3 d\tau_3^2 \right)\right]~~.
\ee
The radii of curvature of the AdS$_4$ and of $\IP^3$ factors are
\be
R^2_{\rm AdS}=\frac{R^3}{4k_{\rm cs}} \ , \ \ \ \ \ 
~~~~~~~~
R^2_{\rm \IP^3}=4R^2_{\rm AdS}\ .
\label{Rads}
\ee
At the world-sheet tree level, the relation between the radius of
curvature and the gauge theory 't~Hooft coupling arises from simply
matching the charges of the supergravity soliton describing the
relevant stack of branes, and to leading order in the strong coupling
expansion one finds \cite{ABJM}
\be
R^2_{\rm AdS}=\sqrt{{\bar\lambda}}~~.
\label{Rstring}
\ee
Due to the non-maximal supersymmetry of the space this relation may,
in principle, receive world-sheet quantum corrections (see footnote
\ref{fnt:corrections} below).\foot{We shall ignore this possibility
here. One way to determine if this relation is modified would be to
study possible renormalization of 3-point functions of chiral primary
operators both on the gauge theory and string theory sides.}  We have
used here the notation ${\bar\lambda}$ introduced in \rf{hed} to
maintain a formal similarity with string theory in \AdSS, where the
radius of the space is the 't~Hooft coupling of the dual gauge theory.

While not entering in the interactions of the world-sheet bosons, the
flux fields govern the interactions of the bosons and the
Green-Schwarz fermions. In that context, their tangent space
components are relevant. For the field strengths in (\ref{bkgrnd})
these components read
\be
(F_2)_{\mu\nu}=2\frac{k_{\rm cs}^2}{R^3}\ellJ_{\mu\nu}\ , 
~~~~~~~~
(F_4)_{abcd}={6}\frac{k_{\rm cs}^2}{R^3}\epsilon_{abcd}\ , 
\ee
or 
\be
e^{\phi}(F_2)_{\mu\nu}=\frac{1}{R_{\rm AdS}}\ellJ_{\mu\nu}\ , 
~~~~~~~~
e^{\phi}(F_4)_{abcd}=\frac{3}{R_{\rm AdS}}\epsilon_{abcd} \ . 
\ee
An important property of \adsc\ space, largely similar to that of
\AdSS\ space, is that all relevant tangent space tensors are constant. Indeed,
here $\ellJ$ and $\epsilon$ are numerical tensors with entries $\pm1$
and $0$. They are, respectively, the entries of the K\"ahler form and
of the volume form on unit $\IP^3$ and AdS$_4$.


All classical spinning string solutions with sufficiently few charges
are common between string theory in \AdSS\ and \AdSP, since
AdS$_4\subset\,$AdS$_5$ and, up to a change of radius, a single
isometry direction looks the same in S$^5$ and $\IP^3$
(for a 
discussion of related classical string solutions which excite more fields
in $\IP^3$ see \cite{Chen:2008qq}). Like the
spinning folded string, the circular rotating string is also in this
class of common solutions. They, in fact, excite the same fields,
which makes them ideal to identify potential conceptual differences
between strings in \AdSS\ and \AdSP.

The world-sheet action is
\be
S&=&S_{\rm AdS_4}+S_{\IP^3}\nn\\
 &=&\frac{R^2_{\rm AdS}}{4 \pi}\int \ d\tau \int_0^{2\pi}
d\sigma\ \sqrt{g}g^{ab } \left(
      G_{\mu\nu}^{\rm AdS}\partial_a X^{\mu} \partial_b X^{\nu}
   +4 G_{\mu\nu}^{\IP^3}\partial_a X^{\mu} \partial_b X^{\nu}
                         \right)~.
\label{bose_action}
\ee 
We express the string tension $T= {\sqrt{\vphantom{|}\bar\lambda}\ov 2
\pi}$ in terms of the radius of the AdS space as in the \adss\
case.\foot{\label{fnt:corrections} This relation may, in fact, receive
quantum corrections. Indeed, since the \AdSP\ geometry is not
maximally supersymmetric, it may be corrected at the world-sheet
quantum level. Type IIA supergravity action is known to receive higher
derivative corrections; modifications of the classical geometry arise
from requiring that the geometry solves the modified equations of
motion. Such higher derivative corrections, however, first arise at
order ${\cal O}(\alpha'^3)$, i.e.  they would be suppressed by an
additional factor of $\lambda^{-3/2}$. They are thus of too high an
order to be relevant to the one-loop calculation we will be interested
in here.}  We will be using the conformal gauge and thus take the worldsheet
metric to be flat, $g_{ab}=\eta_{ab}.$

All conserved quantities derived from this action are related to the
corresponding charge densities by factors of the string tension:
\be
(E,\,S,\,J)= \sqrt{\bar\lambda}\,({\cal E},\,{\cal S},\,{\cal J})~~,
\ee
where $({\cal E},\,{\cal S},\,{\cal J})$ are given in terms of the
momenta conjugate to isometry directions from the Lagrangian
\be
L={ 1 \ov 2} 
\eta^{ab} 
\left(G_{\mu\nu}^{\rm AdS}\partial_a X^{\mu} \partial_b X^{\nu}
   +4 G_{\mu\nu}^{\IP^3}\partial_a X^{\mu} \partial_b
   X^{\nu}\right)~~.
\ee
The rotating 
string solution we are interested in lies in an
AdS$_3\times$S$^1$ subspace  of AdS$_4 \times \IP^3$. 
The choice of the circle S$^1\subset \IP^3$
should be such that it corresponds to the BMN vacuum state
chosen on the gauge theory side,  i.e. a gauge-invariant combination
 ${\rm Tr} (Y^1 Y^\dagger_4)^J$ of the scalar field 
 bilinear  $Y^1 Y^\dagger_4$ \ci{gaiotto,ter}. 
 
It is useful to discuss in more detail how 
the  $\IP^3$ coordinates in  (\ref{p3_metric}) are
related to the scalar fields of the dual  gauge theory of \ci{ABJM}. 
It is natural to start with the form of the metric written in 
projective coordinates
(see e.g.   \cite{Hoxha:2000jf}). Given  an 
eight-dimensional flat space $ds^2=dZ^Ad{\bar Z}_A$  with 
 complex coordinates 
$Z^A$ with $A=1,2,3, 4$,  restricting to the 7-sphere  $\sum_A|Z^A|^2=1$, 
choosing $Z^4=e^{i\tau_4}|Z^4|$ and then introducing $\xi^m=Z^m/Z^4$ with 
$m=1,2,3$ one ends up with  the $S^7$ metric written as a circle fibration
over $\IP^3$. Rewriting $\xi^m$ in terms of its  norm and a unit vector $u^m$ as  
$\xi^m=\tan\zeta_1\ u^m$ one may then repeat this construction
recursively. 

The isometric directions of the resulting metric denoted
by $\tau_{1,2,3}$  correspond to the phases of the analogs of $Z^4$
at each step of the recursion, i.e.
 
  $Z^4=e^{i\tau_4}|Z^4|$,\ \ $Z^3=e^{i(\tau_3+\tau_4)}|Z^3|$, \ \ 
$Z^2=e^{i(\tau_2+\tau_3+\tau_4)}|Z^2|$, \ \ $Z^1=e^{i(\tau_1+\tau_2+\tau_3+\tau_4)}|Z^1|$.

\noindent
To identify the spinor representation of the
$SO(6)\subset SO(8)$ R-symmetry of the gauge theory (the scalars 
$Y^A$ are transforming as a spinor) 
 let us  define a new set of angles 
\be
\tau_1=\varphi_2-\varphi_1~,~~
\tau_2=\varphi_3-\varphi_2~,~~
\tau_3=\varphi_2+\varphi_1~,~~
\tau_4=\tau_0+\varphi_4~,\ \ \ \varphi_4\equiv 
-\frac{1}{2}(\varphi_1+\varphi_2+\varphi_3) 
\ee
getting 
\be
& &Z^1=e^{i[\tau_0+\frac{1}{2}(+\varphi_3+\varphi_2-\varphi_1)]}|Z^1|~,~~ \ \ \ \
Z^2=e^{i[\tau_0+\frac{1}{2}(+\varphi_3-\varphi_2+\varphi_1)]}|Z^2|~,\nn\\
& &Z^3=e^{i[\tau_0+\frac{1}{2}(-\varphi_3+\varphi_2+\varphi_1)]}|Z^3|~,~~\ \ \ \ 
Z^4=e^{i[\tau_0+\frac{1}{2}(-\varphi_3-\varphi_2-\varphi_1)]}|Z^4|\ .
\ee
Observing  that the shift by  $\tau_0$  does not affect the $\IP^3$
coordinates, the homogeneous coordinates $Z^A$ are in one-to-one
correspondence with the four gauge theory scalar fields $Y^A$ in the spinor
representation of $SO(6)$, provided  one identifies the three Cartan 
generators of $SO(6)$ as  represented by shifts of $\varphi_{1,2,3}$,
i.e.
%
$ J_i=-i\frac{\partial}{\partial\varphi_i}.$
 Then 
\be
J_1(Z^A)= (-\ha,\ha,\ha , - \ha ) \ , \ \ \ \  
J_2(Z^A)= (\ha,-\ha,\ha , - \ha ) \ , \ \ \ \ 
J_3(Z^A)= (\ha,\ha,-\ha , - \ha ) \ .
\ee
Thus, the $SO(6)$ charges of the operator $\Tr[(Y^1Y^\dagger_4)^J]$
are matched by the charge of the product of $J$ bilinears,
$[Z^1(Z^4)^\dagger]^J$.
Since $Z^1(Z^4)^\dagger = e^{ i (\varphi_2+\varphi_3)} |Z^1||Z^4|$
this means that $\varphi_2+\varphi_3$ should have nontrivial
background.  Then $J_1 (Z^1(Z^4)^\dagger) =0,\ \ J_2
([Z^1(Z^4)^\dagger]^J) = J_3 ([Z^1(Z^4)^\dagger]^J) = J.  $ To
guarantee that the vacuum contains no other fields it is necessary to
require that $\varphi_2-\varphi_3$ and $\varphi_1$ have trivial
background.\footnote{Then in terms of homogeneous coordinates $Z^A$,
the phases of only $Z^1$ and $Z^4$ will be nonvanishing which is
consistent with having a bilinear combination $(Y^1Y^\dagger_4)$ in
the spin chain vacuum \ci{mz}. }
In terms of the original coordinates $\tau_{1,2,3}$ this translates
into 
\be \tau_2=0\ ,\ \ \ \ \ \tau_1=\tau_3 
\ , 
\ee
which may be realized if the
coordinates $\zeta_i$ in   \rf{p3_metric} take the background values
\be \la{sola}
{\bar \zeta}_1=\frac{\pi}{4},\qquad 
{\bar \zeta}_2=\frac{\pi}{2},\qquad 
{\bar \zeta}_3=\frac{\pi}{2}\  .
\ee
Then the  relevant part of the full 10-d metric becomes 
\be
ds^2=R_{\rm AdS}^2\left[-\cosh^2\rho\ dt^2 +d\rho^2
+\sinh^2\rho\ (d\theta^2+\sin^2\theta \,d\phi^2) +
d{(\varphi_2+\varphi_3)}^2\right] \ . 
\ee
The values of the remaining coordinates on the solution of \cite{PTT}
here are (${\boldsymbol{\sigma}}= (\tau, \sigma)$)
\be
{\bar t}=\kappa\tau={\hat {\rm n}}\cdot {\boldsymbol{\sigma}}\ ,
\ \ \ \bar \rho=\rho_*\ , && \ \    \bar \theta = {\pi \ov 2} \ , \ \ \ 
~~
{\bar \phi}={\rm w}\tau+k\sigma={\tilde {\rm n}}\cdot
{\boldsymbol{\sigma}}\ , \\
\bar \varphi_1=0\ ,\ \ \ \ \
{\bar {\varphi}}_2&=&{\bar {\varphi}}_3=\frac{1}{2}(\omega\tau +
m\sigma) = \frac{1}{2} {\rm m}\cdot {\boldsymbol{\sigma}} \ , 
\label{sol}
\ee
where the constant vectors are 
%
\be {\hat {\rm n}}=(\kappa, 0)\ , \ \ \
\ \ \ \ {\tilde {\rm n}}=({\rm w},k)\ , \ \ \ \ \ {\rm m}=(\omega, m)
\ . 
\ee 
Here $k$ and also  $m$ are arbitrary integers  ($\s$-coordinate is $2 \pi$ periodic). Indeed, 
 as one can show  by considering 
the flat space limit of the metric   \rf{p3_metric}, 
the combinations of angles 
$ \tau_3=\vp_2 + \vp_1  $,\  $ \tau_2+\tau_3= \vp_3 + \vp_1  $ 
and $ \tau_1+\tau_2+\tau_3 =\vp_2 + \vp_3 $ 
should  have  $2 \pi$ periodicity, while each $\varphi_i$ is $\pi$-periodic. 

%
Written in terms of $ \vp\equiv \vp_2 + \vp_3 $ with $\bar \vp = 2
\bar \vp_2 = {\rm m}\cdot {\boldsymbol{\sigma}}$ this solution becomes
the same as in \adss:\footnote{
Let us mention that the definition of R-charges used here is different
from the one used in \cite{mlr}; there the charge $J$ was given by the
momentum conjugate to the field $\vp$ and thus is twice as large as
the R-charges used here.
}
 in particular, the relations between the
parameters following from the equations of motion and the Virasoro
constraints are the same as those in the string theory in \AdSS\ case
(cf. \cite{PTT}):\footnote{These relations imply certain useful
identities between the seven parameters entering the solution; one of
them, which will be useful later in the calculation of the fermionic
characteristic frequencies is \cite{PTT}:
\be
\frac{{\rm r}_1(k\omega -{\rm w}m)}{\sqrt{m^2+{\rm r}_1^2k^2}}
=\frac{\omega}{k{\rm r}_1}\sqrt{m^2+{\rm r}_1^2k^2}~~.
\nonumber
\ee}
\be
{\rm w}^2-(\kappa^2+k^2)=0\ , \ \ \ \ \ \ \ \ \ 
{\rm r}_1^2 {\rm w} k +\omega m &=& 0\cr
-{\rm r}_0^2\kappa^2+{\rm r}_1^2 ({\rm w}^2+k^2)+\omega^2+m^2&=&0~~,
\label{constraints}
\ee
\be {\rm r}_0\equiv \cosh\rho_*\ ,\ \ \ \ \ \ \ \ \ \  ~~{\rm r}_1\equiv \sinh\rho_* \ . \ee
From these constraints one may find, e.g. the expression of $(\kappa,
{\rm r}_1^2, {\rm w})$ in terms of $(m,k,\omega)$. The explicit
relations look rather complicated and not very enlightening; below we
will only need their series expansion in a certain limit.

The charge densities are given by
\be 
{\cal E}=\int^{2\pi}_0  \frac{d\sigma}{2\pi}\ r_0^2  \kappa 
        ={\rm r}_0^2\kappa \   , \quad 
{\cal S}=\int^{2\pi}_{0} \frac{d\sigma}{2\pi}\ {r_1^2}{\rm w}
        = {\rm r}_1^2 {\rm w} \ ,\qquad 
{\cal J}_2={\cal J}_3=\int^{2\pi}_{0} \frac{d\sigma}{2\pi} \  \omega
        =\omega,
\label{charges}
\ee
so that  the classical energy, spin and 
the charges under the second and third Cartan generators 
of $SO(6)$ are 
\be
E_0 =\sqrt{\bar\lambda}\, r_0^2 \kappa\ , 
\qquad  \quad  
S=\sqrt{\bar\lambda}\,{\rm r}_1^2 {\rm w}\ , 
\qquad \quad 
J\equiv J_2=J_3=\sqrt{\bar\lambda}\,\omega~,
\ee
while the Virasoro constraint in \rf{constraints} implies that
%
\be k S +  J m =0  \ . \la{vira} \ee
As already mentioned these are exactly the same as the \adss\ case. 

Similarly to the \AdSS\ case, a (technically) useful limit is that of
large spin ${\cal S} $ and large angular momentum $ {\cal J}$ with
their ratio $u$ (and also $k$) held fixed, i.e.
\be
{\cal S},\ \,{\cal J}\rightarrow\infty\ , \ \ \ \ \ \ \ \ \ \ \
u=-\frac{m}{k}=\frac{\cal S}{ {\cal J}}=\frac{S}{ J}={\rm fixed} \ . \label{scaling}
\ee
In this limit it is possible to solve perturbatively the constraints
(\ref{constraints})
\be
 \kappa &=& \omega + \frac{k^2 }{ 2 \omega^2}u (2+u)
         - \frac{k^4}{8\omega^3}u(4 +12 u +8 u^2 +u^3)
         +  {\cal O}\left(\frac{1}{\omega^5}\right)\ , \nn\\
 {\rm r}_1^2 &=& u - \frac{k^2}{2\omega^2}  u (1+u)^2  
         + \frac{k^4}{8\omega^4}  u (1+u)^2(3+10u+3u^2) 
         + {\cal O}\left(\frac{1}{\omega^6}\right)\ , \nn\\
{\rm w} &=& \omega-\frac{k^2}{2\omega}(1+u)^2
         -\frac{k^4}{8\omega^3}(1+u)^2(1+6u+u^2)
         +{\cal O}\left(\frac{1}{\omega^5}\right)\ . 
\ee
Using these expressions, the expansion of the classical energy at
large $\J$ and thus large angular momentum
$J=\sqrt{{\vphantom{{{}^{|^A}}}}\bar\lambda} {\cal
J}=\sqrt{{\vphantom{{{}^{|^A}}}}\bar\lambda}\, \omega $ is given by
\be
E_0&=&  S +  J 
    + \frac{ {\bar\lambda} }{2J} k^2u(1+u) 
    -  \frac{{\bar \lambda}^2}{8J^3}k^4 u(1+u)(1+3u+u^2)\nn\\
& & +\ \frac{{\bar \lambda}^3}{16J^5}k^6 u(1+u) (1+7u+13u^2+7u^3+u^4)
+{\cal O}\left(\frac{1}{J^7}\right)\ . 
\label{E0}
\ee
This result is essentially the same as in the \AdSS\ case \rf{ops}
provided one identifies the two tensions, i.e. $\sqrt{
\lambda_{{\rm AdS_5}}}\to \sqrt{{\vphantom{{{}^{|^A}}}} \bar\lambda}
$. 

A formally alternative    prescription  that also 
relates  the \adss\ and \adsc\ results for the classical string  energy,  
is  \  (i) to replace $\sqrt{ \lambda_{{\rm AdS_5}}} \to 
2 \sqrt{{\vphantom{{{}^{|^A}}}} {\bar\lambda}}$ in \rf{ops}, 
and \ \ (ii) to replace   $E$, $S$ and $J$ in \adss\ result by $2E$, $2S$ and $2J$
(i.e. $S \to 2S$, $J \to 2J$ and add an extra overall  1/2 factor in the energy). 
At this classical level this is obviously equivalent to no rescaling at all:
changing the string tension by 2  is compensated by rescaling of charges by 2
so that classical parameters remain the same.


As we shall see, it is a generalization (with 
$2\sqrt{ {\bar\lambda}} \to 2\bar h (\bar \lambda)$) of 
  the second  prescription that will 
actually extend to the quantum level.  This  should  not be too surprising 
since the  two quantum string theories  appear to  be  quite different. 

It is an analog of this generalized second prescription that was
proposed, from the Bethe Ansatz perspective, in \ci{gv} as a relation
between the universal scaling functions (or leading terms in the
folded string energies) in AdS/CFT$_4$ and AdS/CFT$_3$ cases.
As we shall demonstrate below, quite remarkably, 
this prescription
applies also  to the non-trivial quantum circular string case as well
as to the generalized folded string case with non-zero orbital
momentum $J$ discussed in \ci{mlr,AAB}.

\section{The spectrum of quadratic fluctuations}
\label{sec:Fluctuations}
\subsection{Bosons}

It is not hard to expand the string action (\ref{bose_action}) around
the solution (\ref{sol}). This is, however, largely unnecessary since,
using the close connection to the circular string solution in
AdS$_5\times$S$^5$, we can quickly write down the characteristic
frequencies for the bosonic fluctuations.  The six fluctuations from
the $\IP^3$ split into one massless, four ``light" degrees of freedom
\be
p_0=\sqrt{p_1^2+\frac{1}{4}(\omega^2- m^2)}\ , 
\ee
and one ``heavy" fluctuation
\be
p_0=\sqrt{p_1^2+(\omega^2- m^2)}~.
\ee
From the AdS space one finds one massless degree of freedom, one
massive one
\be
 p_0=\sqrt{p_1^2+\kappa^2}\ , 
\ee
and two fluctuations whose dispersion relation is given by the roots
of the quartic equation
\be\la{boo}
(p_0^2-p_1^2)^2+ 4 r_1^2 \kappa^2 p_0^2
-4\left(1+r_1^2\right)\left(\sqrt{\kappa^2+k^2}\ p_0-k p_1\right)^2=0\ .
\ee
As in \AdSS\ \ci{PTT}, the explicit solution to this equation looks
complicated, but may be constructed perturbatively in the limit
(\ref{scaling}). Furthermore, one can determine the appropriate signs
with which these modes contribute to the energy correction in a
similar fashion to \ci{PTT} by considering the behavior of the
frequencies at large $\omega$.

\subsection{Fermions}

Since the solution has non-zero angular momentum along $\IP^3$, the
spectrum of fermionic fluctuations could be constructed by starting
with the coset superstring action of \cite{AF,Stef}. This is, however,
not necessary here; instead, we will use the standard form of the
quadratic part of the $\kappa$-symmetric Green-Schwarz action
\be
L_{2F}=i(\eta^{ab}\delta^{IJ}-\epsilon^{ab}s^{IJ})
{\bar\theta}^Ie\llap/{}_a D^{JK}_b\theta^K~~ .
\label{fermi_action}
\ee
Here $s^{IJ}={\rm diag}(1,-1)$ and $e_a^A=\partial_a X^M E_M^A$, where
$X$ denote generic coordinates and $E_M^A$ is the vielbein.  In the
string frame the type IIA covariant derivative is (see e.g.
\cite{Hassan_Tduality} for a choice of field variables with nice
transformation properties under T-duality)
\be
D^{JK}_a&=&\left(\partial_a+\frac{1}{4}\partial_a X^M
\omega_M{}^{AB}\Gamma_{AB}\right)\delta^{JK} -\frac{1}{8}\partial_a
X^M E_M^A H_{ABC}\Gamma^{BC}(\sigma_3)^{JK}\cr
&+&\frac{1}{8}e^{\phi}\left[
  F_{(0)}(\sigma_1)^{JK}
+ F\llap/{}_{(2)}(i\sigma_2)^{JK}
+ F\llap/{}_{(4)}( \sigma_1)^{JK} \right] e\llap/{}_a
\ee
The spin connection components  in the AdS directions are:
\be
&&\omega^{01}=-\omega^{10}=\sinh\rho\; dt~, ~~~~~~~~~~~~~~~~~~\,
\omega^{21}=-\omega^{12}=\cosh\rho\; d\theta\ , \cr
&&\omega^{31}=-\omega^{13}=\cosh\rho\ \sin\theta \  d\phi\ , ~~~~~~~~~~
\omega^{32}=-\omega^{23}=\cos\theta \; d\phi\ . 
\ee
The spin connection components along $\IP^3$ are more complicated but,
due to our choice of coordinates, they will not be needed in this
leading-order calculation.


To find the fermionic spectrum  we 
evaluate the fermionic action  (\ref{fermi_action}) on the background
solution (\ref{sol}) and then impose a gauge-fixing condition which is
adapted to the resulting kinetic operator:  one needs to make
sure that the resulting operator is invertible. 

The features of the resulting kinetic operator may be exposed through
a series of constant field redefinitions which map the background
vielbein to a scalar multiple of a single Dirac matrix. Then, after
combining the two type IIA fermions of opposite chirality into a
single unconstrained 32-component spinor $\psi$ and also using the
symmetry properties of the ten-dimensional Dirac matrices, the
fermionic kinetic operator (\ref{fermi_action}) becomes manifestly
proportional to the projector
\be
{\cal P}_+=\frac{1}{2}\left(1+\Gamma_0\Gamma_3\Gamma_{-1}\right)~~.
\ee
The natural $\kappa$-symmetry gauge  then is 
\be
{\cal P}_+\psi=\psi~~.
\ee
We relegate the details of this calculation, as well as the
construction of the eigenvalues of the resulting quadratic operator,
to appendix \ref{app:fermions} and record here only the
conclusions. The spectrum contains four different frequencies, each
being doubly-degenerate. Two such pairs have frequencies
\be
(p_0)_{\pm 12}=\pm \frac{{\rm r}_0^2 k\kappa m}{2(m^2+{\rm r}_1^2k^2)
}
+\sqrt{(p_1\pm b)^2+(\omega^2+k^2{\rm r}_1^2)}\ , ~~~~~~~~~
b=-\frac{\kappa m}{{\rm w}}\frac{{\rm w}^2-\omega^2}{2(m^2+{\rm
r}_1^2k^2)
}
\ee
while the frequencies of the other two pairs are solutions of the
equation
\be\la{kpq}
(p_0^2-p_1^2)^2+  {\rm r}_1^2 \kappa^2  p_0^2
-\left(1+{\rm r}_1^2\right)\left(\sqrt{\kappa^2+k^2}\ p_0-k p_1\right)^2=0~~.
\ee
The latter equation may be mapped to a similar one in the bosonic case
\rf{boo} by replacing $k$ and $\kappa$ with $2k$ and $2\kappa$ (or
equivalently by replacing $p_0$ and $p_1$ with $\tfrac{1}{2}p_0,\
\tfrac{1}{2}p_1$).  The constant shifts of several of the fermionic
frequencies are similar to those found in the case of the folded
string and, in fact, even for the short and fast BMN string. They may
be removed (at least at the level of the quadratic action) by a
further time-dependent redefinition of the fermions.
We will not,  however,  do this here: as is easily seen, they simply
cancel among themselves when we consider the sum over all frequencies
and so these constant shifts do not contribute to the one-loop
correction to the energy.

Let us note that the superconformal algebra 
supercharges -- and thus the Green-Schwarz fermions -- transform in
the ${\bf 6}_0\oplus {\bf 1}_2\oplus {\bf 1}_{-2}$ of the $SU(4)$
R-symmetry group. In the presence of the rotating string background
the $SU(4)\simeq SO(6)$ breaks to $SO(4)=SU(2)\times SU(2)$.  This
breaking pattern and the spectrum listed above are consistent if we
associate the degenerate fermion pairs with the self-dual and
anti-self-dual representations of $SO(4)$ or pairs of singlets related
by charge conjugation.
One may test this in the BMN limit: the fermion spectrum splits in two
sets of four modes of equal masses; since the R-symmetry group is
$SO(4)$ and two modes are R-symmetry singlets, it follows that in this
limit the spectrum decomposes as ${\bf 4}_0\oplus {\bf 1}_0\oplus {\bf
1}_{0}\oplus {\bf 1}_2\oplus {\bf 1}_{-2}$.

\section{One-loop correction to the string  energy}

The expression for the correction to the string energy can be found by
summing the frequencies over all flavours and mode numbers
\be
E_1=E_1^{(0)}+{\bar E}_1\ , 
\ee
where $E_1^{(0)}$
is the contribution of the zero modes
and ${\bar E}_1$ involves the infinite sum over all non-zero modes (we
set $p_1\equiv n=0, \pm 1, ...$)
\be
\la{eee}   E_1^{(0)}={1\ov 2\kappa}e(0) \ , \ \ \ \ \ \ \ \ \  \ \ \ \ \ \  
{\bar E}_1=-\frac{1}{2 \kappa}\,e(0)+\frac{1}{2 \kappa} \sum_{n=-\infty}^{\infty} e(n)~~.
\ee
The summand $e(n)$ is simply the weighted sum of the bosonic and
fermionic frequencies found  in the previous section:\foot{
The contribution of  two massless degrees of freedom cancels
against the contribution of the diffeomorphism ghosts.}
\be
e(n)&=&   \frac{1}{2}\left[(p_0)^B_{1}+(p_0)^B_{2}-(p_0)^B_{3}-(p_0)^B_{4}
\right]
+\sqrt{n^2+\kappa^2}+ \sqrt{n^2+(\omega^2-k^2 u^2) }\nn\\
& &+\ 4 \sqrt{n^2+\frac{1}{4} (\omega^2-k^2 u^2)}
-2\sqrt{(n-b)^2+(\omega^2+ k^2 r_1^2)}-2\sqrt{(n+b)^2+(\omega^2+ k^2
r_1^2)}
\nn\\
& & -\ \left[(p_0)^F_{1}+(p_0)^F_{2}-(p_0)^F_{3}-(p_0)^F_{4}\right]\ ,  \la{ee}
\ee
where $(p_0)^B_{i}$ and $(p_0)^F_{i}$  stand for solutions of the
quartic equations \rf{boo} and \rf{kpq}.

\bigskip

Before proceeding, let us make few comments about the derivation of \rf{eee}.
This expression for the one-loop correction to the energy of the
rotating string may be arrived at in several different ways. One can use
the expression for the string energy in conformal gauge in terms of
the fluctuation fields derived in appendix A of \ci{FT1} (in that
paper this was in the  context of the
folded spinning string):
\be
E_1=\frac{1}{\kappa}\langle\Psi|H_2|\Psi\rangle\ , 
\ee
with $H_2=\int \tfrac{d\sigma}{2\pi} {\cal H}_2({\tilde t},{\tilde
\phi},\dots)$ being the quadratic worldsheet Hamiltonian corresponding
the fluctuation action at this order.

As here we are  interested only in the one-loop result, this
Hamiltonian approach is sufficient and practical. However, certain
conceptual issues are perhaps clearer in the path-integral
approach. In the AdS$_5 \times $S$^5$ theory where two-loop
calculations have been performed it has been found useful to extract
the correction to the string energy from the sigma model partition
function. It was argued in
\cite{Frolov:2006qe,Roiban:2007jf,Roiban:2007dq} and in greater detail in 
\cite{Roiban:2007ju} that for a homogeneous   string solution like the 
one we consider here $E_1$ may be defined as the one-loop effective
action divided by the two-dimensional time interval. Moreover, the
(quantum-corrected) charges of the background solution are also
determined by the one-loop effective action and, similarly to the
energy, are finite at this order.

In a path integral approach the frequency sum appearing in the
Hamiltonian formalism arises in the process of evaluating the
logarithm of the regularized determinants of the operators of
quadratic fluctuations around the classical solution. Though the final
result is finite, as may be seen by inspecting the large mode-number
behavior of frequencies, each determinant taken separately is
divergent. As a consequence of the path integral approach, all
determinants are regularized in the same way.

An advantage of this approach is that field/fluctuation redefinitions
are systematically accounted for the path integral evaluation of the
effective action or free energy. Such redefinitions (e.g. the ones
equivalent to changing the original coset representative) may
effectively lead to constant shifts in the frequencies of various
modes. While {\it a priori} such shifts may lead to (power-like)
divergences in the free energy, their contribution is, in fact,
canceled exactly by the Jacobian due to the change in the measure of
the path integral and thus it does not change the expression for the
energy shift.

\bigskip

\subsection{Large spin expansion of   one-loop correction to the energy }

While computing exactly the sum over frequencies in \rf{ee} is
difficult, there is one particular region of the parameter space that
is amenable to explicit evaluation: this is the scaling region
\rf{scaling}, i.e. that of large angular momentum $\J$ or large
$\omega$, and large spin $\S$ with the ratio
$u=-\frac{m}{k}=\frac{\cal S}{\cal J}$ (and also $k$) fixed.  As
discussed in \cite{bt} in the context of string theory in \AdSS, in
this limit the sum over modes receives contributions from two distinct
regions:

(I)\ $n\ll \omega$: \ \  here  the sum remains discrete 

(II)\ $n/\omega=x=$fixed: \  here  the sum may be replaced by an
integral over $x$

\noindent
These two regimes are compatible; while each regime exhibits
singularities, it is possible to see that the singular part of  one
regime is captured by the regular part of the other. Thus, the
complete result as an expansion in $1/\omega$ is the sum of the
regular parts of the two regimes, 
\be
 E_1=\frac{1}{2\kappa}\sum_{n=-\infty}^{\infty}e(n)=
\frac{1}{2\kappa}\sum_{n=-\infty}^{\infty} e^{\rm sum}_{\rm reg}(n)
 +  \frac{\omega}{2\kappa} 
\int^{\infty}_{-\infty} \ dx\  e^{\rm int }_{\rm reg}(x)=
E_1^{(0)}+ {\bar E}_1^{\rm even}+{\bar E}_1^{\rm odd}~~.
\label{split_regimes}
\ee
It is an interesting question whether the zero-mode part $E_1^{(0)}$
should be kept separate or whether it effectively belongs to ${\bar
E}_1^{\rm even}$ or ${\bar E}_1^{\rm odd}$.  As we will argue shortly,
it belongs to ${\bar E}_1^{\rm odd}$ part, i.e.  $E_1^{\rm odd}=
E_1^{(0)} +{\bar E}_1^{\rm odd}$.

It is not difficult to solve perturbatively the quartic equations
\rf{boo} and \rf{kpq} and find the most 
non-trivial bosonic and the fermionic frequencies at large 
$\omega$:  
\be
(p_0)^B_{1,3}&=&\frac{p_1}{2\omega}
\left[2 k(1+u)\pm\sqrt{p_1^2+4 k^2 u(1+u)}\right]+{\cal O}\left(\frac{1}{\omega^3}\right)\nn\\
(p_0)^B_{2,4}&=&\pm 2 \omega \pm 
\frac{1}{2\omega}\left[p_1^2\mp2 k p_1(1+u)+2k^2(1+u(3+u))\right]+
{\cal O}\left(\frac{1}{\omega^3}\right)
\\
(p_0)^F_{1,3}&=&\frac{p_1}{\omega}
\left[k(1+u)\pm\sqrt{p_1^2+ k^2 u(1+u)}\right]+{\cal O}\left(\frac{1}{\omega^3}\right)\nn\\
(p_0)^F_{2,4}&=&\pm  \omega \pm 
\frac{1}{\omega}\left[p_1^2\mp k p_1(1+u)+\tfrac{k^2}{2}(1+u(3+u))\right]
+{\cal O}\left(\frac{1}{\omega^3}\right)\ . 
\ee
Then, the summand $e(n)$ in equation \rf{split_regimes} as a function
of the momentum mode number $n$ takes the form
\be
&&e^{\rm sum}(n)=\frac{1}{2 \omega}
\bigg[ n\bigg(3n-4\sqrt{n^2+k^2u(1+u)}+\ \sqrt{n^2+4k^2u(1+u)}\ \bigg) \nn \\
&& \ \ \ \ \ \ \ \ \ \ \ \ \ \ \ \ \ \ \ \ \ \ \ \
- \ k^2(1+u)(1+3 u)\bigg]+{\cal O}\left(\frac{1}{\omega^3}\right).
\label{summand_1_leading}
\ee
The sum over $n$ is singular with a divergence arising from the
constant term which also gives rise to the zero mode piece of the
energy. This occurs at one order lower in the $1/\omega$ expansion
than for the rotating string in AdS$_5\times$S$^5$. Continuing to
higher orders in $1/\omega$ one finds the same splitting into regular
and singular parts, $e^{\rm sum}=e^{\rm sum}_{\rm reg}+e^{\rm
sum}_{\rm sing}$.

The contribution of large mode numbers, $n=\omega x$ with fixed $x$,
may be accounted for by replacing the sum over $n$ with an integral
over $x$. To leading order in the large-$\omega$ expansion the summand
becomes
\be\la{ei}
e^{\rm int}(x)&=& \frac{k^2(1+u)}{2 \omega} 
\left[\frac{1+u(3+2 x^2)}{(1+x^2)^{3/2}}
- 2\frac{1+u(3+8 x^2)}{(1+ 4 x^2)^{3/2}}\right]
+{\cal O}\left(\frac{1}{\omega^3}\right)~~,
\ee
where one can see that $e^{\rm int }_{\rm reg}(n/\omega)=e^{\rm
sum}_{\rm sing}(n)$. It is interesting to note that, while capturing
the singular part of the sum over $e^{\rm sum}(n)$ in equation
(\ref{summand_1_leading}), it also correctly captures the zero-mode
contribution:
\be
e^{\rm int }_{\rm reg}(0)=e^{\rm sum}_{\rm sing}(0)~~.
\ee
Thus, we may simply combine the zero-mode contribution $e^{\rm
sum}(0)$ together with the contribution of large mode numbers.
It is possible to extend the comparison above to $e^{\rm sum}_{\rm
reg}$ and $e^{\rm int }_{\rm sing}$ to higher orders in the $1/\omega$
expansion,
which we carry out explicitly in appendix \ref{sum_vs_integral} and
show that indeed $e^{\rm int }_{\rm sing}(x)=e^{\rm sum}_{\rm
reg}(\omega x)$ to all orders we checked. This is exactly analogous to
the recombination which takes place in \AdSS\ case.

Since the sum is absolutely convergent, the coefficients in the $1/J$
expansion of the discrete part of the correction to the energy may be
computed as formal power series in $k$
\be
{\bar E}_1^{\rm even}&=&\frac{1}{\kappa}\sum_{n=1}^{\infty}
 e^{\rm sum}_{\rm reg}(n)\nn\\[5pt]
&=&
-\frac{{\bar\lambda}
k^4(1+u)^2 u^2}{2^3J^2}
\left(6\zeta(2)-15 k^2u(1+u) \zeta(4)
+\frac{315}{8}k^4u^2(1+u)^2\zeta(6) +\dots \right)\nn\\
& &+\frac{{\bar\lambda}^2
k^6(1+u)^2 u^2}{2^6 J^4} \Big(24(1+2u-u^2)\zeta(2)+{15} k^2 u^2(1+u)(5+13u)\zeta(4)
\nn\\
& &\kern+95pt-\frac{63}{2}k^4u^2(1+u)^2(5+22u+27u^2)\zeta(6)
+\dots\Big)\nn\\
& &-\frac{{\bar\lambda}^3
k^8(1+u)^2 u^2}{2^9J^6}
\Big({48}(3+18u+26u^2+10u^3+7u^4)\zeta(2)
\nn\\
& &\kern+95pt-{60}k^2u^2(1+u)(7+27u+53u^2+49u^3)\zeta(4)\nn\\
& &\kern+95pt+{63}k^4u^2(1+u)^2(5-20u-183u^2-382u^3-264u^4)\zeta(6)
+\dots\Big)\nn\\
& &+\ {\cal O}\left(\frac{1}{J^8}\right).
\label{E1even}
\ee
Using the expression for $e^{\rm int}$ listed in appendix
\ref{sum_vs_integral} to go to higher orders in the $1/J$ expansion,
the continuum contribution to the energy reads:
\be
E_1^{\rm odd}&=&\frac{\omega}{2\kappa}  \int^{\infty}_{-\infty}\  \  dx\  
e^{\rm int }_{\rm reg}(x)\nn\\[5pt]
&=&-\ \frac{\, {\bar\lambda}^{1/2} k^2 }{J} \ \ln 2\ u(1+u)
+\frac{ \,{\bar\lambda}^{3/2} k^4}{2J^3}\  \ln 2 \ u(1+u)(1+3 u+u^2) \nn\\
& & -\ \frac{\,  {\bar\lambda}^{5/2} k^6 }{8J^5}u(1+u)
\Big[3(1+7u+13u^2+7u^3+u^4)\ln 2-\frac{4}{3}u^2(1+u)^2\Big]\nn\\
& &
 +\ {\cal O}\left(\frac{1}{J^7}\right)~~.
\label{E1odd}
\ee
While anticipated by the existence of divergences in the discrete
contribution to leading nontrivial order, the appearance of such low
odd powers of $1/J$ with ``non-analytic'' factors of $\bar \lambda$
may at first look surprising. It is possible to test numerically that
the expressions above are indeed accurate (see appendix
\ref{numerics1}).

\subsection{Relation to the energy of the circular rotating string in \AdSS}
Motivated by the similarity of the classical solution we started with to the one in \adss\ 
 and also by the fact that the proposed Bethe Ansatz of \ci{gv} 
has a structure similar to that of the  AdS/CFT$_4$ case 
let us now compare the result for $E_1$ to the corresponding expression 
in \adss\ string theory.\foot{Note that the fluctuation frequencies
in the two theories are not directly related (corresponding to the two 
superficially quite different 2d quantum theories), 
but their respective sums representing $E_1$'s  happen to be 
similar as we describe below.}

Collecting the results of the previous section, the total one-loop
corrected energy of the circular rotating string in \AdSP\  is
\be\la{kq}
E=E_0+ E_1= E_0 +  {\bar E}_1^{\rm even}+E_1^{\rm odd}~~,
\ee
with $E_0$, ${\bar E}_1^{\rm even}$ and $E_1^{\rm odd}$ are given by
equations (\ref{E0}), (\ref{E1even}) and (\ref{E1odd}), respectively.

From equation (\ref{E0}) we note that the classical energy of the
circular rotating string in the scaling (large spin \rf{scaling})
limit is a series in inverse odd powers of the angular momentum
$J$. One may then contemplate that $E_0$ and $E_1^{\rm odd}$ might
naturally combine together. This is indeed the case as we may write
their sum as
\be
E_0+E_1^{\rm odd}&=&S +  J +\frac{{\bar h}^2(\bar \lambda) k^2}{2J}u(1+u)
-\frac{ {\bar h}^4 (\bar \lambda) k^4}{ 8J^3} u(1+u)(1+3u+u^2)\nn\\
& &\ \ + \ \frac{  {\bar h}^6(\bar \lambda)k^6}{16 J^5}u (1+u)
(1+7u+13u^2+7u^3+u^4)\nn\\
& &\ \ + \ \frac{ {\bar h}^5(\bar \lambda)k^6}{6 J^5}u^3 (1+u)^3+{\cal
O}\left(\frac{1}{J^7}\right)\ .
\label{dd}
\ee
Here we  introduced the function 
\be\la{ahh}
{\bar h}(\bar \lambda)=\sqrt{\bar\lambda }-{\ln 2}
+{\cal O}\left(\frac{1}{\sqrt{\bar\lambda}}\right) \ . 
\ee
The powers of ${\bar h}(\bar \lambda)$ in equation (\ref{dd}) are
understood to be truncated to the two leading terms in
$1/\sqrt{\bar\lambda}$ expansion except for the last term,
proportional to ${\bar h}^5(\bar \lambda)/J^5$ which is understood to
be truncated to the leading term. This is indeed the correct
prescription, as the ${\bar\lambda}$ dependence of the next-to-leading
term identifies it as a two-loop correction.

For comparison, let us recall the analogous part of the expression for
the one-loop energy of the same circular rotating string in the \AdSS\
case \cite{PTT,bt} (see \rf{ops},\rf{lps} and the discussion in the
introduction)
\be
(E_0+E_1^{\rm odd})_{_{\rm AdS_5\times S^5}}&=&
J+S+\frac{\lambda k^2}{2J}u(1+u)
-\frac{\lambda^2 k^4}{8 J^3} u(1+u)(1+3u+u^2)\nn\\
& &+ \ \frac{\lambda^3 k^6}{16J^5}u (1+u)
(1+7u+13u^2+7u^3+u^4)\nn\\
& &+\ \frac{\lambda^{5/2} k^6}{3 J^5}u^3 (1+u)^3+{\cal
O}\left(\frac{1}{J^7}\right) \ . 
\label{Ad}
\ee
We then observe that the \adsc\ expression \rf{dd} can be obtained
from the \adss\ one \rf{Ad} by the prescription mentioned earlier at
the end of section 2:
\be 
E_{_{\rm AdS_4\times \IP^3}}^{\rm odd} (S,J,k;\ \sqrt{\bar \l}\ )\  
=\ {1 \ov 2}  E_{_{\rm AdS_5 \times S^5}}^{\rm odd} 
(2S,2J,k;\ 2 \bar h ({\bar \l})\ ) \ , 
\la{pro} 
\ee
with the function ${\bar h}({\bar\lambda})$ given by \rf{ahh}.  It is
important to note that the replacement $\sqrt{\lambda}\mapsto 2 {\bar
h}(\bar \lambda)= 2 \sqrt{{{\bar \lambda}}\vphantom{{}^{{}^{A}}}} - 2
\log 2 + ...$ is to be implemented {\it after} the energy is expressed
in terms of the conserved charges $(S,J)$ (which are also the
parameters on the gauge theory side).

Notice that what selects  between the simple  replacement $\l \to h(\bar \lambda)$
with no change to the  charges and the prescription \rf{pro} 
(which were equivalent at the classical level) is the matching of the 
last ``quantum phase'' term in \rf{dd} and in \rf{Ad}.

\

As was already mentioned in the introduction (see eq.\rf{bmn} and
discussion below it) the one-loop renormalization of the leading
``analytic'' terms in the \adsc\ string energy implies that the BMN
spectral relation here gets a one-loop renormalization, i.e.  the
function \rf{ahh} should be identified with the function ${\bar
h}(\bar \lambda)$ entering the magnon dispersion relation
(cf. \rf{dis},\rf{hed})
\be
\epsilon(p)=\frac{1}{2}\sqrt{1+\frac{4}{\pi^2}
{\bar h}^2(\bar \lambda)\sin^2\frac{p}{2}}~~.
\label{man}
\ee
Notice that \rf{man}  is related to the familiar  \adss\  expression
$\epsilon(p)=\sqrt{1+\frac{\l }{\pi^2}\sin^2\frac{p}{2}}$ by the same 
prescription \rf{pro}  (cf. also \rf{bmn}).

One useful way to understand the relation between the renormalization
of the magnon dispersion relation and the above function ${\bar
h}(\bar \lambda)$ is to consider the analog of the effective
Landau-Lifshitz (LL) model description of the large $J$ limit as was
done in the \adss\ case in \ci{kru,mtt}.  The LL model may be viewed
as an effective 2d field theory which describes the ``fast string'' or
large $J$ expansion on both the string and spin chain side and thus
interpolates between the two descriptions. Considering for
illustrative purposes the analog of the $SU(2)$ sector action
parameterized by a unit 3-vector ${\vec n}$ the corresponding LL action
is $S= J \int dt \int { d \sigma \over 2 \pi} \ L$ where \ci{mtt}
\be
L&=&C({\vec n})\cdot \partial_0 {\vec n} -  
      {\vec n}\bigg[
      \sqrt{1-\,{4{\bar h}^2(\bar \lambda)\ov J^2} {\partial_1^2}}-1\bigg]
      {\vec n}-{ a(\bar\lambda)\ov J^4} (\partial {\vec n})^4\cr
      & &\ \ \ \ \ \ \ \ \ \ \ -\ {1 \ov J^6}  \left[
             b_1(\bar \lambda)(\partial_1 {\vec n})^2(\partial^2_1 {\vec n})^2
            +b_2(\bar \lambda)(\partial_1 {\vec n}\cdot\partial^2_1 {\vec n})^2
            +b_3(\bar \lambda)(\partial_1 {\vec n})^6\right]+\dots \ . 
\la{nk}
\ee
In general, $\bar h$, $a$, $b_i$, etc., are interpolating functions
parameterizing this low-energy effective action. In the
\adss\ case the first three functions are simple: $2 \bar h\to
\sqrt{ \lambda}$, $a\to { 3 \ov 128} { \lambda}^2$, $b_1\to -{
7 \ov 4} { \lambda}^3$. The functions $b_2,b_3$ are non-trivial,
having the same ${r \l}^3$ behaviour at weak and strong coupling
but with different numerical coefficients (reflecting the ``3-loop
disagreement'').

All of these  functions  are expected to be nontrivial in the 
present \adsc\ case.\foot{In particular, 
due to the structure of perturbation theory in the ${\cal N}=6$ CS
theory the function $a(\bar\lambda)$ should start  at weak coupling 
with a 4-loop $\bar \l^4$ term.}
By comparing the energy of the rotating string as described by the LL
action with the explicit string theory computations one observes that
the $u\to 0$ limit of (\ref{dd}) should be essentially captured by the
leading quadratic in $\vec n$ terms in \rf{nk}, thus identifying $\bar
h$ in \rf{ahh} with the function that governs the magnon dispersion
relation (\ref{man}). 


\

Remarkably, the same prescription \rf{pro} also relates the folded
string energies in \adss\ and \adsc.  Indeed, ignoring first the
$J$-dependence, starting with the \adss\ one-loop result \ci{FT1}
\be 
E_{_{\rm AdS_5 \times S^5}}=S + {1 \ov \pi}( \sqrt \lambda - 3 \log 2
) \ln S + ...\ , 
\ee 
and making the replacements in \rf{pro} one finds (for $S \gg 1$) 
\be 
E_{_{\rm AdS_4\times \IP^3}}=S + 
{1 \ov 2\pi}[ 2 \bar h^2 (\bar \l)   - 3 \log 2 ]  \ln S  + ...\ . 
\ee
Using the expression \rf{ahh} for $\bar h$ found here we end up with 
\be 
E_{_{\rm AdS_4\times \IP^3}}=
S + {1 \ov \pi}( {\sqrt{\bar \l} }    
- {5 \ov 2}  \log 2 )  \ln S  + ...\ , \ee  
which is the expression found by direct string computation 
in \ci{mlr,AAB,CK}.
Moreover,  by including the dependence on $J$ (in the limit of large $\S$ with 
${ \J \ov \ln \S}$ fixed)  one finds that the equations
\rf{pro}, \rf{ahh} directly relate the  \adss\ result of \ci{Frolov:2006qe}
to the one in the \adsc\ case as found in \ci{mlr,AAB}.
This provides a nontrivial  consistency check  between  currently
available one-loop results in the \adsc\ superstring.


%
 It should be noted, however, that the prescription \rf{pro} is so far
 rather  heuristic (or empirical, on the string theory side) and need
 not a priori apply to the whole expression for the one-loop string
 correction.\foot{It does seem to apply to the ``non-analytic'' part
 of the one-loop correction, which comes from the ``integral'' term in
 the one-loop calculation and is not sensitive to the compactness of
 the worldsheet; this is also the only term that determines the
 leading one-loop shift in the folded string case. It is, in principle,
 possible that a different prescription is necessary to map the
 ``analytic'' part of the one-loop correction to the corresponding
 \AdSS result.
}

\

Returning 
to the circular string solution, the relation between the equation
(\ref{dd}) and the corresponding result in \AdSS\ (\ref{Ad}) via the
equation \rf{pro} suggests to compare also the terms containing even
powers of $1/J$ in (\ref{E1even}) with the analogous terms in the
\AdSS\ case.
The part of the \adss\ one-loop energy which is proportional to the
even inverse powers of the $S^5 $ angular momentum is
(cf. \cite{PTT})\foot{In the \adss\ case the zero-mode contribution to
the energy is also accounted for by the contribution of large mode
numbers. 
Note also that here we are assuming that one can interchange summation 
with expansion in $1/J$, but otherwise there is no regularization ambiguity 
(as would be  present in the Landau-Lifshitz model approach) as we start with the
full  UV finite expression for the sum.}
\foot{Here we record only the regular contributions to the discrete sum from the \adss\ case. The 
divergent contribution starts at order $\tfrac{1}{J^6}$ and corresponds to the non-analytic 
contribution coming from the dressing phase at order $\tfrac{1}{J^5}$
\ci{bt}. Keeping only the regular 
contributions is equivalent to evaluating 
the summation using the zeta-function regularization as was
done in \ci{szz}.}
\be
({\bar E}_1^{\rm even})_{\rm AdS_5\times S^5}&=&
\frac{1}{\kappa}\sum_{n=1}^\infty 
e^{\rm sum}_{\rm reg,\, AdS_5\times S^5}(n)\nn\\[5pt]
&=&-\ \frac{{\lambda}k^4(1+u)^2 u^2}{2^2J^2}
\left(4\zeta(2)-8 k^2u(1+u) \zeta(4)
+20k^4u^2(1+u)^2\zeta(6) +\dots \right)\nn\\
& &+\ \frac{{\lambda}^2k^4(1+u)^2 u^2}{2^5J^4}
\Big(16k^2(1+2u-u^2)\zeta(2)+{8} k^4 u^2(1+u)(5+13u)\zeta(4)
\nn\\
& &\kern+120pt- \ 16k^6u^2(1+u)^2(5+22u+27u^2)\zeta(6)
+\dots\Big)\nn\\
& &- \ \frac{{ \lambda}^3k^4(1+u)^2 u^2}{2^8J^6}
\Big(32k^4(3+18u+26u^2+10u^3+7u^4)\zeta(2)
\nn\\
& &\kern+10pt\ \ \ \ - \ 32k^6u^2(1+u)(7+27u+53u^2+49u^3)\zeta(4)\nn\\
& &\kern+10pt\ \ \ \ +\  32k^8u^2(1+u)^2(5-20u-183u^2-382u^3-264u^4)\zeta(6)
+\dots\Big)\nn\\
& &+\ {\cal O}\left(\frac{1}{J^8}\right). \la{sa}
\ee
Comparing this with equation (\ref{E1even}) we note that, while 
not exactly the same, 
 the two expressions may be mapped into each other by again 
replacing $\sqrt{\lambda}\mapsto 2{\bar h}(\bar \lambda)$,
$S \to 2S$, $J\to 2J$   (i.e. $u \to u$)  and  $E\to 2E$
as in \rf{pro}
but in addition also by replacing $\zeta(n)$ in the \AdSS\ result
 \rf{sa} by
\be\la{saa} 
\zeta(n)\mapsto  2\Big(1-\frac{1}{2^n}\Big)\zeta(n)~~.
\ee
This modification of the $\zeta$-constants in the \AdSS\ calculation 
may be formally interpreted as replacing the sum over even mode
numbers $n$  in \rf{E1even} by a sum over odd mode numbers,   
\be
\sum_n \omega_n=\sum_n \omega_{2n}+\sum_n \omega_{2n+1}~\mapsto~
2\sum_n \omega_{2n + 1}~~.
\ee 

\newpage


\subsection{Comments on comparison to the Bethe Ansatz proposal}

In a finite two-dimensional quantum field theory, loop corrections to
the conserved charges (such as the target space energy) of classical
solitons may be  computed using the standard perturbative approach, either
in the Hamiltonian or in the path integral setting.  If this
two-dimensional theory is 
 dual, through gauge/string duality, to some planar gauge theory, then
 the target space energies obtained this way in an acceptable (in the
 sense defined in the introduction) regularization scheme should yield
 the strong coupling expansion of the anomalous dimensions of certain
 gauge theory operators.

If this two-dimensional theory is also integrable, then 
its semiclassical states can be described using the algebraic curve
techniques
\ci{kmmz}, which also determines the fluctuation frequencies \ci{dvi,bfr,gv1}
near the solitonic solutions and thus, effectively, the 1-loop
corrections to their charges. Furthermore, there may exist a set of
(discrete) Bethe equations that should provide the exact description
of quantum corrections to all loop orders.
The results of the algebraic curve approach   and the Bethe Ansatz  approach 
should of course agree  with 
 the results found by  the  direct worldsheet computations, 
 and this should be, in fact, a test of the validity of the algebraic
curve and Bethe Ansatz approaches. 
%

In the Bethe Ansatz approach one solves directly the algebraic
(``discrete'') Bethe equations and thus no choice of regularization is
required. Such a choice is, however,
required in the algebraic curve approach, where, similarly to the
worldsheet calculation, one finds the frequencies of small
fluctuations near a soliton from an algebraic curve and then uses the
standard quantum-mechanical prescription to evaluate the one-loop
correction by computing the sum of frequencies (weighted by $(-1)^{F}$
where $F$ is the fermion number). Since the all-order Bethe Ansatz
construction is based on a ``discretization'' \ci{afs} of the
classical (integral) Bethe equations \ci{kmmz} and since their
solution requires no regularization, it follows that a special choice
of regularization is required in the algebraic curve calculation to
reproduce the results of the Bethe Ansatz calculation.

This  is the case for strings in \AdSS, where the Bethe equations and
the worldsheet calculation yield the same result which is matched by
the algebraic curve calculation \ci{bfr,gv1}
provided one chooses a natural
regularization which accounts for certain constant shifts in the
space-like momenta of fluctuations 
and essentially amounts to introducing 
different cutoffs for various partial frequency sums (cf. \ci{gv1,fpt,btz,szz}).


For string theory in \AdSP\ the same three strategies are, in
principle, also available. In particular, for the circular rotating
string we have already the worldsheet results obtained in the
previous subsection. One may also consider the implications of the
algebraic curve approach \ci{gv2,gv_new} and of the Bethe Ansatz
equations proposed in \cite{gv} to describe the corresponding set of
gauge theory operators with one spin and one R-charge.\foot{It is
worth pointing out that this rank one sector, in fact, captures only
{\it some} of the $sl(2)$ sector solutions, namely those with odd
Bethe mode number. The other solutions mix with the other sectors,
requiring the use of the complete set of nested Bethe equations.}
Below we shall  only  briefly  comment  on the corresponding 
 solution to the Bethe Ansatz  equations  and its  comparison 
with the worldsheet approach.\foot{More details  
and comparison with the algebraic curve approach should appear in  Ref.
 \cite{gv_new}.} 
 The relevant Bethe Ansatz equations are given by:
\be
\left(\frac{x_l^+}{x_l^-}\right)^{2J}=-\prod_{j\ne l=1}^S
\frac{u_l-u_j+i}{u_l-u_j-i}
\left(\frac{x_l^--x_j^+}{x_l^+-x_j^-}\right)^2
\sigma_{\rm BES}^2(u_l,u_j)~~,
\label{GV_be}
\ee
where\foot{ The charge $J$ used here in the Bethe equations is the
same as the angular momentum $J$ used in our worldsheet calculation
and the gauge theory R-charge $J$ which in turn is half the spin-chain
length.}
\be
x^\pm+\frac{1}{x^\pm}=\frac{1}{h(\lambda)}\left(u\pm\frac{i}{2}\right)\ , 
\ee
and  $h(\lambda)$ is the interpolating function in the magnon dispersion relation (cf. 
\rf{dis},\rf{hed}).  Here  $\sigma_{\rm BES}$ is the same dressing
phase as in the \ 
context \ci{bes}, but with $\sqrt{\lambda}$ replaced
by $4\pi h(\lambda)$ in the appropriate way \ci{gv}. We will consider
a class of solutions of \rf{GV_be}, with the one-cut solution of
\ci{KZ} particularly in mind, vis-\`a-vis those of the analogous
$sl(2)$ Bethe equation in \AdSS\ to which it has a great degree of
similarity.
The total energy of  the  solutions is given by 
\be
E-J=2i h(\lambda)\sum_{l=1}^S\left(\frac{1}{x^+_l}-\frac{1}{x^-_l}\right)\ , 
\ee
or in terms of the magnon momenta
\be
E-J=\sum_{l=1}^S\sqrt{1+16h^2(\lambda)\sin^2\frac{p_l}{2}}\ , \la{eey}
\ee
and the zero-momentum condition is
\be
\Big[\prod_{l=1}^S\left(\frac{x_l^+}{x_l^-}\right)\Big]^2=1~~.
\label{zero_mom}
\ee
The absence of the factor of $1/2$ in the expression for the energy
\rf{eey} and the square in \rf{zero_mom} are due to the identification
of the $u_4$ and $u_{\bar 4}$ roots \cite{gv}.

As was mentioned in \cite{gv}, the only differences between the
equations above and the analogous ones in \AdSS\ are the replacement of
the square-root of the 't~Hooft coupling $\lambda$ of ${\cal N}=4$ SYM
with $4\pi h(\lambda)=2{\bar h}({\bar\lambda})$, the different relation between
the R-charge and the spin-chain length and the existence of
an additional minus sign on the right-hand-side of (\ref{GV_be}). This
additional sign is  like a familiar ``magnetic field'' twist and the corresponding Bethe Ansatz 
equations are also analogous to those that appear in the 
$\beta$-deformed SYM theory \ci{ma,frt}  for a special real
value of the deformation parameter $\beta_d=\frac{1}{2J}$. 
Indeed, the $\beta$-deformed Bethe equations and the 
zero momentum condition  are \cite{br}
\be
e^{-2i\pi\beta_d J} \left(\frac{x_l^+}{x_l^-}\right)^{2J}&=&
\prod_{j\ne l=1}^S
\frac{u_l-u_j+i}{u_l-u_j-i}
\left(\frac{x_l^--x_j^+}{x_l^+-x_j^-}\right)^2
\sigma_{\rm BES}^2(u_l,u_j)~~.
\la{bff}
\ee
This equation becomes the same as equation (\ref{GV_be}) upon choosing
$\beta_d=\frac{1}{2J}.$ \foot{The zero-momentum condition, 
$e^{-2i\pi\beta_d S}\prod_{l=1}^S \frac{x_l^+}{x_l^-}=1,$ is, however,
different from  eq.(\ref{zero_mom}) by a factor of $(-1)^{S/J}$. It is the
consequences of the latter equation which we will discuss here.}
%

\def \td {\tilde}


In   the $\beta$-deformed  context
\ci{frt,br}  the only effect of the phase $\beta_d$  is to
shift the integer number that appears in the logarithm of the Bethe
equations by $\beta_d J$. 
This  may  be seen directly  by
taking the logarithm of equations (\ref{GV_be}) and (\ref{zero_mom}):
\be
2\pi i  \big({\tilde k}+\ha \big) +2 J \ln \frac{x_l^+}{x_l^-}&=&\sum_{j\ne l=1}^S
\ln \bigg[\frac{u_l-u_j+i}{u_l-u_j-i} \left(\frac{x_l^--x_j^+}{x_l^+-x_j^-}\right)^2
\sigma_{\rm BES}^2(u_l,u_j)\bigg] \ , 
\la{lop} \\ 
2\pi i {\tilde m} + 2 \sum_{l=1}^S \ln \frac{x_l^+}{x_l^-} &=& 0 \ , \la{lopp}
\ee
where ${\tilde k}\in \IZ$ is the Bethe mode number and $\tilde m$ is an integer.
Consistency of the equations \rf{lop} and \rf{lopp} implies that
\be
 {\tilde m} J+ ({\tilde k}+\ha)S=0~~.
\label{log_constraint}
\ee
Compared to the corresponding equations in the \AdSS\ case there are 
five changes: 

(1) here  $4 \pi h(\l)$ is in place of $\sqrt{\l}$;

(2) here the spin chain length  is $2J$  not $J$; 

(3) here the energy of a solution is doubled (due to the double number
of excitations);

(4) the BPS condition at vanishing coupling requires that spin $S$ be
doubled;\foot{That is, items $(3)$ and $(4)$ imply that 
$E=S+J+\dots~~\to~~2E=2S+2J+\dots$}

(5) here  
${\tilde k}+\ha$ is in place  of $\td k$  (due to the additional 
minus sign in the equation (\ref{GV_be})).

\noindent
The square in the equation \rf{zero_mom} and together with 
the doubled number of excitations (point (3) above) imply that no
change in ${\tilde m}$ is necessary.
With these identifications \rf{log_constraint} is formally the same as
the usual constraint in the \AdSS\ case: ${\tilde m}( 2 J) +
({\tilde k}+\ha)(2S)=0\ \to\ m J + k S =0 $. In the case of the circular
(rational) solution we are interested in, ${m}$ and ${ k}$ are,
respectively, the S$^5$ and the AdS$_5$ winding numbers.

The solution of these Bethe equations in the strong coupling
limit, to the leading and subleading order, proceeds as in the \adss\
case
\cite{btz,szz} (see \cite{bt,hl} for the inclusion of the one-loop
corrections to the dressing phase).
To obtain the solution of the Bethe equations (\ref{GV_be})
from that of the \AdSS\  Bethe equations with length $J$ and parameters 
$k, m, S$ with $m J + k S=0$  one is then   to  make the following formal 
 replacements as implied by the above discussion:

\

$\!\!\!\!$
(1) $ \sqrt \l \to 4 \pi h(\l)$;\ \ (2) $J \to 2 J $; \ \ 
(3) $S \to 2 S $; \ \ (4) $E \to 2 E $;  \ \ 
(5) $k \to \td k + \ha$;\ \ (6) $m \to \td m.$

\

\noindent
Comparing now  the solution of the above AdS/CFT$_3$  Bethe equations 
to the  classical \AdSP\ string solution discussed in  section 2 we are led
to the following identification:

(i) \   ${\tilde k}+\ha  \ \to  k $,  where $k$ is the 
AdS$_4$ winding number;\ \ 

(ii) \ $\tilde m\ \to  m $, where   $m$ is the 
$ \IP^3$ winding number.

\noindent
Note that  \rf{log_constraint} becomes then equivalent to the Virasoro constraint in 
\rf{vira}. 

The above replacements  
  reproduce
the energy of the classical rotating string in \AdSP\ from the energy
of the classical rotating string in \AdSS.
%
Moreover, 
 keeping the next to leading term in  $h(\lambda)$
 reproduces all the non-analytic terms in $(E_0+E_1^{\rm odd})_{\rm
 AdS_4\times \IP^3}$, those related to the classical energy of the
 string as well as those related to the corrections from the one-loop
 phase.

In addition, the various identifications of parameters which relate
the Bethe Ansatz energy with the result of the worldsheet calculation
for the circular rotating string also lead to the correct map for the
folded spinning string and the universal scaling function, as may be
seen by inspecting the Bethe Ansatz solution in \cite{KC}. 

%

Unfortunately, the same cannot be said about the relation between 
the analytic 1-loop  terms 
$(E_1^{\rm
even})_{\rm AdS_5\times S^5}$ in \rf{sa} 
and $(E_1^{\rm even})_{\rm AdS_4\times
\IP^3}$  in \rf{E1even}.
Using the above  identifications in $(E_1^{\rm even})_{\rm AdS_5\times
S^5}$, it appears that an additional formal replacement
for the $\zeta$-constants is needed. 
This is exactly the same replacement
described earlier, \rf{saa}, and  has the same interpretation of replacing the sum
over even modes by an additional sum over odd modes.
%
Such a replacement, however, seems unjustified on the basis of the
Bethe equations (\ref{GV_be})--(\ref{zero_mom}). 

\

To summarize, we have found that the conjectured all-loop Bethe Ansatz
\ci{gv} reproduces the general structure of the \AdSP\ superstring
calculation.

In particular, the worldsheet approach predicts that the function
$h(\l)$ that should be used in the Bethe Ansatz proposal of \ci{gv}
should be given by \rf{hah},\rf{hh} (or, equivalently \rf{ahh}). This
conclusion (as well as the confirmation that the strong-coupling limit
of the phase in the Bethe Ansatz should be, indeed, the same as in
\adss\ case)
 is not sensitive to the compactness of the worldsheet  $\s$  direction. 

However, the matching of the analytic $1/ J^{2n}$ terms in the string
1-loop energy (whose coefficients {\it are} sensitive to the
compactness of the $\sigma$ direction) is not immediately clear.
It might be that we are missing some subtlety in the identification of
the circular string configuration as a Bethe Ansatz solution, or that
some details of the Bethe Ansatz proposal of \ci{gv} still need to be
adjusted. Further analysis is required to settle these issues.

\


\bigskip
\subsection*{Acknowledgments}

We are  grateful to  G. Arutyunov,  N. Beisert, S. Frolov, S. Sch\"afer-Nameki, K. Zarembo 
and especially N. Gromov  and P. Vieira
 for numerous   useful discussions    and helpful  comments on the manuscript. 
The work of RR is supported by the Department of Energy OJI award 
DE-FG02-90ER40577, the National Science Foundation under grant
PHY-0608114 and the A.P. Sloan Foundation.

\newpage

\appendix

\section{
Details of  calculation of the
fermionic spectrum \label{app:fermions}}

Our starting point is the action in \rf{fermi_action}.
We will analyze separately the geometric and the flux part of the
covariant derivative.
The action has constant coefficients and the kinetic operator may be 
extracted directly. For the purposes of analytic calculations,  however,  
it is useful to first perform certain field redefinitions.

The background value of the slashed vielbein is 
\be
e\llap/{}_a={\hat {\rm n}}_a \Gamma_0
           +{\tilde {\rm n}}_a\Gamma_3
           +{\rm m}_a\Gamma_9
\ee
Here as in \rf{sol}  \  ${\hat {\rm n}}=(\kappa, 0)$, ${\tilde {\rm n}}=({\rm w},k)$ 
and ${\rm m}=(\omega, m)$.
A sequence of two rotations with constant coefficients 
\be
S=S_{39}S_{09}~,~~&&~~~
S_{39}=\cos p +\sin p \Gamma_{39}~,~~~~~~~~
S_{09}=\cosh q +\sinh q \Gamma_{09}\cr
&&\sin 2p=\frac{m}{\sqrt{m^2+k^2{\rm r}_1^2}}
~,~~~~~~~~~~~
\sinh 2q=-\frac{\omega}{k {\rm r}_1}
\ee 
transform $e\llap/{}_0$and $e\llap/{}_0$ into two Dirac matrices.
Then the transformed value of the slashed vielbein becomes:
\be
S^{-1}e\llap/{}_0 S=\sqrt{m^2+k^2{\rm r}_1^2}\;\Gamma_0
~,~~~~~~~~~~
S^{-1}e\llap/{}_1 S=\sqrt{m^2+k^2{\rm r}_1^2}\;\Gamma_3\ . 
\label{vielb_transf}
\ee
Due to the choice of isometry direction $\varphi_3$ in the
construction of the circular rotating string solution, all components
of the spin connection along $\IP^3$ vanish when evaluated on the
background. Thus, the geometric part of the (transformed) covariant 
derivatives is as in \cite{PTT}:
\be
D_a&\equiv &\partial_a+\frac{1}{4}\omega_a{}^{AB}\Gamma_{AB}\ , \cr
D^S_0=S^{-1}D_0S&=&\partial_0
-\frac{{\rm r}_0{\rm r}_1k^2}{
2\sqrt{m^2+{\rm r}_1^2k^2}}\Gamma_{01}
-\frac{{\rm r}_0{\rm r}_1kw}{2\sqrt{m^2+{\rm r}_1^2k^2}
}\Gamma_{13}
+\frac{\kappa m}{{\rm w}}\frac{{\rm w}^2-\omega^2}{2(m^2+{\rm
r}_1^2k^2)
}\Gamma_{19}, 
\cr
D^S_1=S^{-1}D_1S&=&\partial_1 
+\frac{{\rm r}_0}{{\rm r}_1}\frac{m\omega}{2\sqrt{m^2+{\rm r}_1^2k^2}
}\Gamma_{01}
-\frac{{\rm r}_0{\rm r}_1k^2}{2\sqrt{m^2+{\rm r}_1^2k^2}
}\Gamma_{13}
+\frac{{\rm r}_0^2 k\kappa m}{2(m^2+{\rm r}_1^2k^2)
}\Gamma_{19}
\ee
In the type IIA theory the fermions $\theta^{1,2}$ are chiral and 
of opposite chirality
\be
\Gamma_{-1}\theta^1=\theta^1,\ \ \ \ \ \ \ \ \ 
~\Gamma_{-1}\theta^2=-\theta^2~~.
\ee
They may be combined into a single 32-component unconstrained spinor 
$\psi=\theta^1+\theta^2$. Then, 
\be
s^{1J}\theta^J+s^{2J}\theta^J=\Gamma_{-1}\psi~~.
\ee
With these observations the geometric part of the action for the
fermionic quadratic fluctuations may be written as
\be
(\eta^{ab}\delta^{IJ}-\epsilon^{ab}s^{IJ})
{\bar\theta}^Ie\llap/{}_aD_b\theta^J&=&{\bar\psi}e\llap/{}_a
D_b(\eta^{ab}\id-\epsilon^{ab}\Gamma_{-1})\psi\cr
&=&
-{\bar\psi}e\llap/{}_0D_0 \psi
+{\bar\psi}e\llap/{}_1D_1 \psi
-{\bar\psi}e\llap/{}_0D_1\Gamma_{-1}\psi
+{\bar\psi}e\llap/{}_1D_0\Gamma_{-1}\psi\cr
&=&
-{\bar\psi'}\Gamma_0(1+\Gamma_{03}\Gamma_{-1})D^S_0\psi'
+{\bar\psi'}\Gamma_3(1+\Gamma_{03}\Gamma_{-1})D^S_1\psi'
\ee
where $\psi'=(m^2+k^2{\rm r}_1^2)^{1/4}S^{-1}\psi$. Opening the
parenthesis one finds without difficulty that the terms in $D_i^S$ which
do not commute with $(1+\Gamma_{03}\Gamma_{-1})$ cancel out either
among themselves or because of the constraint ${\bar \psi}\Gamma_A\psi=0$. 
One is finally left with
\be
(\eta^{ab}\delta^{IJ}-\epsilon^{ab}s^{IJ})
{\bar\theta}^Ie\llap/{}_aD_b\theta^J &=&
{\bar\psi'}(-\Gamma_0D_0+\Gamma_3D_1)(1+\Gamma_{03}\Gamma_{-1})\psi'
\ee

Let us focus next on the flux-dependent terms in the super-covariant derivative. 
Their contribution to the action \rf{fermi_action} as well as the precise expressions
for the slashed fluxes are
\be
(\eta^{ab}\delta^{IJ}-\epsilon^{ab}s^{IJ})
{\bar\theta}^Ie\llap/{}_a
\left[F\llap/{}_2(i\sigma^2)^{JK}+F\llap/{}_4(\sigma_1)^{JK}\right]
e\llap/{}_b\theta^K
&=&
{\bar\psi}e\llap/{}_a\left[-F\llap/{}_2\Gamma_{-1}+F\llap/{}_4\right]e\llap/{}_b
(\eta^{ab}\id+\epsilon^{ab}\Gamma_{-1})\psi
\no \\
F\llap/{}_2=2(\Gamma_{45}-\Gamma_{67}+\Gamma_{89})\ , 
~~~~~~~& &~~~~~~~~~
F\llap/{}_4=6\Gamma_{0123} \ . 
\ee
To simplify this expression we next 
split $S^{-1}(-F\llap/{}_2\Gamma_{-1} +F\llap/{}_4)S$ into a sum of terms
\be
S^{-1}(-F\llap/{}_2\Gamma_{-1} +F\llap/{}_4)S=
{\cal F}+{\cal F}_{03}+{\cal F}_0+{\cal F}_3\ , 
\ee
where ${\cal F}$ commutes with $\Gamma_0$ and $\Gamma_3$, ${\cal F}_{i}$ 
anticommutes with $\Gamma_i$ and ${\cal F}_{ij}$ anticommutes with 
$\Gamma_{ij}$. These properties are sufficient to show that 
${\cal F}_{0}$ and ${\cal F}_{3}$ cancel out and that the only relevant 
terms  will be ${\cal F}$ and ${\cal F}_{03}$ whose expressions are
\be
{\cal F}&=&-2(\Gamma_{45}-\Gamma_{67})\Gamma_{-1}-2\cosh 2q\cos 2p\Gamma_{89}\Gamma_{-1}\ , \cr
{\cal F}_{03}&=&6\cosh 2q\cos 2p\Gamma_{0123}
\ee 
Using (\ref{vielb_transf}) one may rewrite the flux term in the fermionic action as 
\be
&&{\bar\psi}e\llap/{}_a\left[-F\llap/{}_2\Gamma_{-1}+F\llap/{}_4\right]e\llap/{}_b
(\eta^{ab}\id+\epsilon^{ab}\Gamma_{-1})\psi\cr
&&\ \ = \sqrt{m^2+k^2{\rm r}_1^2}{\bar\psi}'
\left[{\cal F}-{\cal F}_{03}\right](\id+\Gamma_{03}\Gamma_{-1})\psi'\cr
&&\ \ =-\sqrt{m^2+k^2{\rm r}_1^2}{\bar\psi}'
\left[2\left((\Gamma_{45}-\Gamma_{67}) 
+ \cosh 2q\cos 2p\Gamma_{89}\right)\Gamma_{-1}\right.
\cr
&&~~~~~~~~~~~~~~~~~~~~~~~~~~~~~\left.
+6\cosh 2q\cos 2p\Gamma_{0123}\right](\id+\Gamma_{03}\Gamma_{-1})\psi'\ , 
\ee
Then, the complete fermionic quadratic Lagrangian is 
\be
{\cal L}&=&i(\eta^{ab}\delta^{IJ}-\epsilon^{ab}s^{IJ})
{\bar\theta}^Ie\llap/{}_a D^{JK}_b\theta^K=
i{\bar\psi}K\psi'\ , 
\nonumber\\[10pt]
K&=&
\bigg\{ 2(-\Gamma_0D_0^S+\Gamma_3D_1^S)
-\frac{1}{4}\sqrt{m^2+k^2{\rm r}_1^2}\Big[~
6\cosh 2q\cos 2p\Gamma_{0123}
\\
&&\qquad\qquad\qquad\qquad\qquad
+2\left(\vphantom{\big|}
(\Gamma_{45}-\Gamma_{67}) + \cosh 2q\cos
2p\Gamma_{89}\right)\Gamma_{-1}
\Big]\bigg\} {\cal P}_+\ , 
\nonumber
\ee
where 
\be
{\cal P}_+=\frac{1}{2}(\id+\Gamma_{03}\Gamma_{-1})~~.
\ee
The presence of this projector in the quadratic Lagrangian is an indication of
 the $\kappa$-symmetry of the action. A natural$\kappa$-symmetry gauge choice is 
 then that none of the 
components of $\psi$ lie in the orthogonal subspace of ${\cal P}_+$.

Next, we  need to find the frequencies of the fermionic modes described by this 
Lagrangian. To this end it is useful to consider a general operator of  which 
$K$ is a special case. Such an operator is:
\be
{\cal K}=&-&ip_0\Gamma_0+ip_1\Gamma_3 +a\Gamma_{013}+b\Gamma_{019}+c\Gamma_{139}\cr
 &+&6A\Gamma_{0123}+2B\Gamma_{89}\Gamma_{-1}+2C(\Gamma_{45}-\Gamma_{67})\Gamma_{-1}
\ee
where 
\be
\begin{array}{lcl}
a=0\ ,  &~~~~~~~~& 
\displaystyle{A=-\frac{1}{8}\sqrt{\omega^2+k^2{\rm r}_1^2}}\ , \\[8pt]
\displaystyle{b=-\frac{\kappa m}{{\rm w}}\frac{{\rm w}^2-\omega^2}
{2(m^2+{\rm r}_1^2k^2)}}\ , 
&~~~~~~~~& 
\displaystyle{B=-\frac{1}{8}\sqrt{\omega^2+k^2{\rm r}_1^2}}\ , \\[8pt]
\displaystyle{c=-\frac{{\rm r}_0^2 k\kappa m}{2(m^2+{\rm r}_1^2k^2)}}\ , 
&~~~~~~~~&
\displaystyle{C=-\frac{1}{8}\sqrt{m^2+k^2{\rm r}_1^2}} \ . 
\end{array}
\label{coefficients}
\ee
To evaluate the eigenvalues and enforce $\kappa$ gauge fixing ${\cal P}_+\psi = \psi$, 
 let us  find  the eigenvalues
of
\be
K_g={\cal P}^T_+\Gamma_0{\cal K}{\cal P}_+ \ . 
\ee
The factor of $\Gamma_0$ implies that $p_0$ may be extracted from  the zeros of 
the characteristic polynomial. 

To identify the zeros  it is useful to note that the operator $K$ commutes
with $\Gamma_{4567}$ and that the projectors $P_\pm=\ha (1\pm
\Gamma_{4567})$ commute with the $\kappa$-symmetry projector ${\cal P}_+$. 
Then one may  make a further split 
\be
K_{g+}=P_+^TK_gP_+~, ~~~~~~~~~~~~~~~K_{g-}=P_-^TK_gP_-
\ee
The characteristic polynomials for these operators may be found without
difficulty.

The one for $K_{g+}$ implies that $p_0$ is determined by the equation
\be
[-(p_0+c)^2+(p_1-b)^2+4(3A+B)^2]^2[-(p_0-c)^2+(p_1+b)^2+4(3A+B)^2]^2=0 \ , 
\ee
from which one should keep the positive frequencies. The factorized
form means  that there are two doubly-degenerate modes with the frequencies:
\be
(p_0)_{\pm 12}=\pm c+\sqrt{(p_1\pm b)^2+4(3A+B)^2} \ . 
\ee
It is worth noting that these correspond to the ``heavy" fermions. If
one reduces the solution to the case of the BMN string the mass of
these fermions is   twice that of the ``lighter" fluctuations. 

The
frequencies of those lighter modes are determined by the
characteristic polynomial of $K_{g-}$, i.e.  are the roots of
the following quartic polynomial:
\be
\left[(-p_0^2+p_1^2)^2-2 p_0^2 C_{++++}-2 p_1^2 C_{+-+-}
-8 \,b\,c\,p_0p_1+C_{++--}^2\right]^2=0\ , 
\ee
where
\be
C_{+\alpha\beta\gamma}=b^2+4\alpha(3A-B)^2+\beta c^2-16\gamma C^2\ . 
\ee
As for $K_{g+}$, there are two doubly-degenerate modes; upon using the
expressions (\ref{coefficients}) for the constants appearing above,
one finds the equation \rf{kpq} quoted in the text.

\section{Details of   comparison of the fixed and
large mode number contributions to the one-loop string energy \label{sum_vs_integral}}
In this appendix we record the regular and singular terms in the
one-loop frequency sum in the large $\omega$ limit in the discrete and
continuous regimes (see section 4.1) 
\be
e^{\rm sum}=e^{\rm sum}_{\rm reg}+e^{\rm sum}_{\rm sing}\ , 
~~~~~~~~~~~
e^{\rm int}=e^{\rm int}_{\rm reg}+e^{\rm int}_{\rm sing}\ , 
\ee
and compare their structures. The part of the summand $e^{\rm
sum}_{\rm sing}$ that gives rise to a singular contribution in the discrete regime is
\be
e^{\rm sum}_{\rm sing}(n)&=&\!\!
\frac{1}{2 \omega}
\left(-k^2(1+u)(1+3 u)\right)\nn\\
&+&\!\!
\frac{1}{4\omega^3}\left(7k^2(1+u)(3+5u)n^2
+\frac{1}{8} k^4(1+u(44+u(86+(28-15u)u))) \right)\nn\\
&+&\!\!
\frac{1}{16\omega^5}\left(-93k^2(1+u)(5+7u)n^4
-\frac{1}{4}k^4(1+u)(375+u(2509+u(3157+687u)))n^2\right.\nn\\
& &\!\!
\kern+20pt\left. -\frac{1}{16}k^6(1+u)
(1+u(257+u(1134+u(1006+u(65+33u)))))+...\right)\nn\\
&&\!\!
+\ {\cal O}\left(\frac{1}{\omega^7}\right)~~.
\ee
The part of the integrand $e^{\rm int}_{\rm reg}(n)$ that leads to a
regular contribution in the continuum regime is
\be
e^{\rm int}_{\rm reg}(x)
&=&\frac{1}{2\omega}\left( -k^2(1+u)(1+3u)+\frac{7}{2}k^2 (1+u)(3+5 u)x^2
-\frac{93}{8}k^2(1+u)(5+7u)x^4+...\right)\nn\\
&+&\frac{1}{32\omega^3}\Big(k^4(1+u(44+u(86+(28-15u)u)))
\nn\\ & &\kern+140pt
-\frac{1}{2}k^4(1+u)(375+u(2509+u(3157+687u)))x^2\Big)\nn\\
& -&\frac{1}{256\omega^5} \Big(k^6(1+u)
(1+u(257+u(1134+u(1006+u(65+33u)))))+...\Big)\nn\\
& & +\ {\cal O}\left(\frac{1}{\omega^7}\right)\ . 
 \ee
By inspection,  it is not hard to notice that
\be
e^{\rm sum}_{\rm sing}(n)=e^{\rm int}_{\rm reg}({n\ov \omega}) \ , 
\ee
which shows that the regular part in the continuum regime correctly
captures the apparent singularities in the large $\omega$ limit of the
discrete regime.

Similarly, the singular part in the continuum regime 
$e^{\rm int}_{\rm sing}(x)$ and 
the regular part in the discrete regime
$e^{\rm sum}_{\rm reg}(n)$ are
\be
e^{\rm int }_{\rm sing}(x)&=&-\frac{3k^4u^2(1+u)^2}{4x^2\omega^3}+\frac{1}{8 \omega^5}\left(\frac{15k^6u^3(1+u)^3}{x^4}-\frac{3k^6u^2(1+u)^2(-1+2u^2)}{ x^2}\right)\nn\\
 & &+\ {\cal O}\left(\frac{1}{\omega^7}\right)\ , 
\ee
and
\be
\label{eqn:sum_freq_expansion}
e^{\rm sum}_{\rm reg}(n)
&=&\frac{1}{2 \omega}\left(- \frac{3}{2n^2}k^4u^2(1+u)^2
+\frac{15}{4n^4}k^6u^3(1+u)^3+...\right)\nn\\
&+ &\frac{1}{4\omega^3}\left( -\frac{3}{2n^2} k^6 u^2(1+u)^2(-1+2u^2)) 
+ \frac{15}{16 n^4} k^8 u^4 (1+u)^3(13+17u)+... \right)\nn\\
&&+\ {\cal O}\left(\frac{1}{\omega^5}\right)~~,
\ee
respectively. Again, it is not hard to see that 
\be
e^{\rm int }_{\rm sing}(x)=e^{\rm sum}_{\rm reg}(\omega x)~,
\ee
implying that the regular part of the discrete regime correctly
describes the singular part in the continuum regime. 

These
observations parallel those in \AdSS\ made in \cite{bt}. As in that
case, they imply that the one-loop correction to the energy of the
circular rotating string is given by the equation (\ref{split_regimes}).

\section{ Higher orders in the $1/\omega$ expansion of $e^{\rm int}(x)$}

In this appendix we include the expression of $e^{\rm int}$ 
(whose leading order was quoted in \rf{ei}) to higher orders.
%
%
\be
\label{eqn:integrand_freq_x}
e^{\rm int}(x)&=& \frac{k^2(1+u)}{2 \omega^2} \left(\frac{1+u(3+2 x^2)}{(1+x^2)^{3/2}}
- 2\frac{1+u(3+8 x^2)}{(1+ 4 x^2)^{3/2}}\right)\nn\\
& &\kern-20pt- \ \frac{k^4(1+u)}{32\omega^4 x^2} \Big[\frac{1}{(1+ x^2)^{7/2}}
\left(32u^2(1+u)+(7+u(77+u(221+135u)))x^2\right.
\nn\\& &\left.\qquad+4(-7+u(-7+u(29+21u)))x^4+16u(1+u(3+u))x^6+16 u(1+u)x^8\right)\nn\\
& &\qquad -\frac{8}{(1+4x^2)^{7/2}}\left(u^2(1+u)+(1+3u(5+u(11+5u)))x^2\right.\nn\\
& &\qquad\left.+8(-1+3u)(2+u(4+u))x^4+64u(2+3u)x^6+256u(1+u)x^8\right)\Big]\nn\\
& &\kern-20pt+\ \frac{k^6(1+u)}{256\omega^6 x^4} \Big[\frac{1}{(1+ x^2)^{11/2}}
\left(512u^3(1+u)^2+128u^2(1+u)(1+22u+20u^2)x^2\right.
\nn\\& &+(31+u(735+u(3570+u(10418+u(12447+4991u)))))x^4
\nn\\& &  +4(-93+u(-596+u(-907+u(373+u(1412+707 u)))))x^6
\nn\\& &+8(31+u(93+u(254+u(358+u(201+71 u)))))x^8
\nn\\& &+32 u(28+u(132+u(146+u(40+ u))))x^{10}
\nn\\& &+64 u(1+u)(6+u(26+9u))x^{12}
\nn\\& &\left.+32 u(1+u)(3+u)(1+3u)x^{14}\right)\nn\\
& &-\frac{32}{(1+4x^2)^{11/2}} \left(u^3(1+u)^2+u^2(1+u)(1+22u+20u)x^2\right.\nn\\
& &+(1+u(31+u(147+u(357+u(391+157u)))))x^4\nn\\
& &+4(-12+u(-52+u(-9+u(137+u(179+91u))))))x^6\nn\\
& &+16(8+u(64+u(232+u(240+u(67+21u)))))x^8\nn\\
& &+128u(32+u(140+u(142+u(32+u))))x^{10}\nn\\
& &+1024u(1+u)(7+26u+8u)x^{12}\nn\\
& &\left.+2048u(1+u)(3+u)(1+3u)x^{14}\right)\Big]\nn\\
& &+\ {\cal O}\left(\frac{1}{\omega^8}\right).
\ee
At each order in $1/\omega$ one notices terms which are singular as $x\rightarrow 0$. 
These are the terms contributing to $e^{\rm int}_{\rm sing}$ quoted in the previous appendix.

\section{ Numerical checks \label{numerics1}}


The fact that the leading term in the large $\omega$ expansion of the one-loop string energy 
is proportional to $\omega^{-1}$ contrasts with what happened in the case of 
the rotating string in AdS$_5\times$S$^5$, whose ``odd'' part starts only at $ 1/\omega^5$ 
order. 
%
A check of  this dependence may be obtained by a numerical
evaluation of the sum in the regime leading to \rf{E1odd}. 
The main complication is related to
the fact that, while the correction to the energy is finite, each of
the sums contributing to it is divergent. These divergences are of two
types: power-like and logarithmic. While one may directly evaluate the
sums with a cutoff, the presence of divergences leads to a quick loss
of numerical accuracy.

This may be  somewhat improved by separating the sum into two sub-sums and
subtracting the divergences in each of them.\footnote{One may in fact go 
even further and subtract the divergences of each frequency sum separately.} 
Concretely, we split the
full sum into two sums -- over the light and heavy modes;
schematically, they are
\be
e(n)^{\rm light}&=&4\times \sqrt{n^2+\frac{1}{4}(\omega^2-k^2u^2)}
-2\times\frac{1}{2}\left((p_0)^F_1+(p_0)^F_2-(p_0)^F_3-(p_0)^F_4\right)\ , 
\cr
e(n)^{heavy}&=&\sqrt{n^2+\kappa^2}+\sqrt{n^2+(\omega^2-k^2u^2)}
+\frac{1}{2}\left((p_0)^B_1+(p_0)^B_2-(p_0)^B_3-(p_0)^B_4\right)\cr
           &&-2\sqrt{(n-b)^2+(\omega^2+k^2 {\rm r}_1^2)}
           -2\sqrt{(n+b)^2+(\omega^2+k^2 {\rm r}_1^2)}\ ,
\ee
where, as before, $p^{B,F}_{1,2,3,4}$ are solutions of the bosonic and fermionic quartic
equations.

Since the subtracted sums are absolutely convergent, one may carry out
this subtraction for each mode separately. The subtracted terms add up
to zero.
In each of them the power-like divergences cancel out. Then, from each
of them we may  subtract the leading term in the large $n$ expansion for
fixed values of the other parameters 
\be
\Delta S_{light}&=&(\omega^2-k^2-m^2)\frac{1}{2n}\ , 
\cr
\Delta S_{heavy}&=&(3(\kappa^2-\omega^2)-m^2-4k^2{\rm r}_1^2)\frac{1}{2n}\ . 
\ee
These subtractions cancel out when the two sums are added together
because of the usual mass sum rule
\be
\sum_i (-)^{F_i}m_i^2=0~~~\Leftrightarrow~~~~
\kappa^2-m^2-\omega^2-2k^2{\rm r}_1^2=0\ , 
\ee
which here appears as a consequence of the Virasoro constraint.

An unfortunate feature of these sums is that they converge somewhat
slowly in their effective parameter which is $n/\omega$. Indeed, since
the leading large $n$ behavior is $\sim n^{-3}$, the corrections are
of the order $\delta S \sim \ha  (\omega/N)^2$. Consequently, for a
sufficiently large $\omega$ which probes the asymptotic behavior of
the sum, the necessary cutoff $N$ is relatively large.

Numerical evaluation with $\omega=10^4$ and an estimated error
$5\times 10^{-3}$ (i.e. a cutoff $N=10^5$) gives 
\be
2\omega E  =-(1.383\pm .005)k^2 u(1+u)+{\cal
O}(\omega^{-1})=-(1.995\pm .01)
\ln 2\; k^2 u(1+u)+{\cal O}(\omega^{-1})
\ee
Clearly, this is consistent with the leading term in the large
$\omega={J\ov \sqrt{\bar \lambda}}$ expansion obtained analytically in
(\ref{E1odd}).

It is possible, though somewhat cumbersome, to perform similar checks for the subleading terms in the $1/J$ expansion.

\section{ Large $J$, large $k$ limit of circular string solution}

It is interesting to study the large $J$, fixed $S$, 
 limit of the solution considered  in the
main text for, as we will see, this limit 
does not seem to follow the same rules for finding the AdS$_4\times\IP^3$ string energies
from their AdS$_5\times$S$^5$ analogues. 
This limit may, however,  be somewhat   exceptional, as it requires scaling  AdS$_4$
winding number to infinity.
 Nonetheless, if for nothing other than completeness, we
decided to mention it here.



We will consider the limit where 
${\cal J}= \omega$ is taken to be large while ${\cal S}$ and $m$
are kept fixed with $\tfrac{m}{\cal S}<0$. The constraints on the parameters in section 2 
then imply that we must also take  the
winding  $k$  to be large. We will use the notation $k=\beta \omega$ where
 $\beta=-\tfrac{m}{\cal S}$. In this limit the parameters of the solution become
 \be
 \kappa&=&\omega-\frac{m^2}{\sqrt{m^2+{\cal S}^2}}+\order{\omega}\nn\\
 r_1^2 &=&\frac{{\cal S}^2}{\omega \sqrt{{\cal S}^2+m^2}}+\order{\omega^2}\nn\\
 {\rm w}&=&\frac{\omega}{\cal S} \sqrt{ {\cal S}^2 +m^2} +\frac{{\cal S} m^2}{{\cal S}^2+m^2}+\order{\omega}.
 \ee
Then  the energy density,  ${\cal E}$, is infinite but as for the BMN string or giant magnon 
the difference 
${\cal E}-{\cal J}$ is finite, and is given simply by 
\be
E- J= \sqrt{S^2+m^2 \bar{\lambda}}\ \ .\la{kl}
\ee
This classical energy is what we would expect from the analogous AdS$_5\times$S$^5$ result found in 
\ci{mtt3} where 
the one-loop correction was also calculated and shown to be zero.  
Based on the replacement rule, \rf{pro}, we would then 
 expect to find a non-vanishing one-loop contribution proportional 
 to $\ln 2$ in AdS$_4\times\IP^3$ geometry (coming from the 
 $\bar \lambda \mapsto 2 \bar h(\bar \lambda)$ 
 replacement in the classical expression \rf{kl} with $\bar h$ given by \rf{ahh}). 
 However, we will see that this is 
not the case -- the one-loop correction found by direct evaluation from string theory 
  actually vanishes. 


It is straightforward to find the fluctuation frequencies about this 
 large-$J$ solution from
the general frequencies calculated in section \ref{sec:Fluctuations}. From the quartic equation \rf{boo} we find
the characteristic frequencies 
\be
(p_0)^B_{1,2}=\sqrt{(p+\beta)^2+1}\pm \sqrt{1+\beta^2}\ ,~~~ 
(p_0)^B_{3,4}=-\sqrt{(p-\beta)^2+1}\mp \sqrt{1+\beta^2}\ .
\ee
We should note here that we have rescaled the worldsheet coordinate so
that the string has infinite length, scaling like $\omega$.  Thus the
worldsheet momenta, $p$, are now continuous.  From the remaining
bosonic fluctuation frequencies we have six free massive modes -- two
with mass $1$ and four with mass $1/2$. For the fermions we find four
with frequencies calculated from the quartic equation \rf{kpq}
\be
(p_0)^F_{1,2}=\frac{1}{2}\left( \sqrt{(2p+\beta)^2+1}\pm \sqrt{1+\beta^2}\right)\ ,
~~
(p_0)^F_{3,4}=\frac{1}{2}\left(-\sqrt{(2p-\beta)^2+1}\mp \sqrt{1+\beta^2}\right)
\ee
while the remaining four fermions have frequencies
 \be
(p_0)^F_{5,6}=\sqrt{(p+\ha {\beta})^2+1}\pm\frac{1}{2}\sqrt{1+\beta^2},\qquad 
(p_0)^F_{7,8}=\sqrt{(p-\ha {\beta})^2+1}\mp\frac{1}{2}\sqrt{1+\beta^2}. 
\ee
We can now straightforwardly calculate the sum over frequencies which to leading order
in $\omega$ can be replaced by an integral. 
\be
E_1\sim \int_0^\infty dp & &\kern-15pt\Big[2\sqrt{1+p^2}+2\sqrt{1+4 p^2}+\sqrt{1+(p-\beta)^2}
+\sqrt{1+(p+\beta)^2} \\
& &-\sqrt{1+(2p-\beta)^2}-\sqrt{4+(2 p-\beta)^2}-\sqrt{1+(2 p+\beta)^2}-\sqrt{4+(2 p+\beta)^2}\Big].\nn
\ee
If we  follow the standard procedure of imposing 
a cut-off, performing the integral and taking the cut-off to infinity 
we find  that  the one-loop 
correction to the energy of the circle string in this limit is zero.

For comparison, if we  follow \ci{grmik}, we can  identify 
the ``light"  and ``heavy" modes as\foot{This can be done
  by taking the $\beta \rightarrow 0$ which can be viewed as taking ${\cal S}\rightarrow0$ but with $m=0$ and  
which corresponds to the BMN string.}
\be
&&p_0^L=\Big\{ 4 \times \sqrt{1/4+ p^2}, 2 \times \sqrt{1/4+(p\pm \beta/2)^2} \Big\},\\
&&p_0^H=\Big\{ 2\times \sqrt{1+ p^2},\sqrt{1+(p\pm\beta)^2},\sqrt{1+(p\pm\beta/2)^2}\Big\},
\ee
 Then we note that the formula \ci{grmik} for the one-loop energy correction 
\be
E_1=\frac{1}{2\kappa}\sum_{n=-\infty}^{\infty} \big[ p_0^H(n)+\frac{1}{2}p_0^L(n/2) \big]
\ee
becomes,  in the limit of large 
${\cal J}=\omega$
(where we again set $n=\omega p$  and replace the sum by an integral),
exactly half of the analogous result in AdS$_5\times$S$^5$ 
which in this case is  also zero.

\newpage


\end{document}